\documentclass[9pt,twocolumn,twoside]{pnas-new}
\templatetype{pnasresearcharticle}

\usepackage{graphicx}
\usepackage{amssymb}
\usepackage{epstopdf}
\usepackage{url}
\usepackage{subfigure}
\usepackage{soul}
\usepackage{mathtools}
\usepackage{color, colortbl}
\usepackage{fancyhdr}
\usepackage{ctable}
\definecolor{Gray}{gray}{0.9}
\usepackage{enumitem}
\usepackage{graphicx}
\usepackage{caption}
\usepackage{array}
\usepackage{booktabs}
\usepackage{makecell}
\usepackage{cellspace}
\usepackage{tabularx}
\usepackage{xfrac}  
\usepackage{float}
\usepackage{nowidow}

\title{Urban highways are barriers to social ties}

\author[a,b,$\dagger$]{Luca Maria Aiello}
\author[a]{Anastassia Vybornova}
\author[c]{Sándor Juhász}
\author[a,b,c,d]{Michael Szell}
\author[e]{Eszter Bokányi}

\affil[a]{IT University of Copenhagen, Copenhagen, 2300, Denmark}
\affil[b]{Pioneer Centre for AI, Copenhagen, 1350, Denmark}
\affil[c]{Complexity Science Hub Vienna, Vienna, 1080, Austria}
\affil[d]{ISI Foundation, Turin, 10126, Italy}
\affil[e]{University of Amsterdam, Amsterdam, 1018WV, The Netherlands}

\leadauthor{Aiello} 

\significancestatement{Highways are physical barriers that cut opportunities for social connections, but the magnitude of this effect has not been quantified. Such quantitative evidence would enable policy-makers to prioritize interventions that reconnect urban communities -- an urgent need in many US cities. Here we relate urban highways in the 50 largest US cities with massive, geolocated online social network data to quantify the decrease in social connectivity associated with urban highways. We find that this barrier effect is strong in all 50 cities, and particularly prominent over shorter distances. We also confirm this effect for highways that are historically associated with racial segregation. Our research demonstrates with unprecedented granularity the long-lasting impact of decades-old infrastructure on society and provides tools for evidence-based remedies.
}

\correspondingauthor{\textsuperscript{$\dagger$}To whom correspondence should be addressed. E-mail: luai@itu.dk}


\dates{This manuscript was compiled on \today}

\begin{abstract}
Urban highways are common, especially in the US, making cities more car-centric. They promise the annihilation of distance but obstruct pedestrian mobility, thus playing a key role in limiting social interactions locally. 
Although this limiting role is widely acknowledged in urban studies, the quantitative relationship between urban highways and social ties is barely tested. 
Here we define a Barrier Score that relates massive, geolocated online social network data to highways in the 50 largest US cities. At the unprecedented granularity of individual social ties, we show that urban highways are associated with decreased social connectivity. 
This barrier effect is especially strong for short distances and consistent with historical cases of highways that were built to purposefully disrupt or isolate Black neighborhoods. 
By combining spatial infrastructure with social tie data, our method adds a new dimension to demographic studies of social segregation. Our study can inform reparative planning for an evidence-based reduction of spatial inequality, and more generally, support a better integration of the social fabric in urban planning.
\end{abstract}

\begin{document}

\maketitle
\thispagestyle{firststyle}
\ifthenelse{\boolean{shortarticle}}{\ifthenelse{\boolean{singlecolumn}}{\abscontentformatted}{\abscontent}}{}

\dropcap{C}ities are hubs of concentrated social capital that can foster diversity and innovation~\cite{jacobs_death_1961,schlapfer_scaling_2014}. However, this potential is threatened by spatial fragmentation through built infrastructure that can separate neighborhoods~\cite{grannis_importance_1998, anciaes2016urban}, exacerbate inequalities~\cite{ananat_wrong_2011,toth_inequality_2021}, and contribute to segregation~\cite{roberto_spatial_2021}. Among various types of barriers fragmenting urban areas, roads designed for motorized traffic are the most ubiquitous, especially highways~\cite{buchanan1963traffic, anciaes2016urban, rothstein_color_2017}. Since the 1960s, urban planners have theorized that high-traffic roads reduce opportunities for creating and maintaining \emph{social ties} across divided neighborhoods~\cite{appleyard_livable_1981}, thus undermining the social cohesion essential for the development of thriving communities. This premise lies at the core of contemporary urban planning research and interventions~\cite{rueda_superblocks_2018, moreno_introducing_2021} that strive to meet the UN's sustainable development goal of ``making cities and human settlements inclusive, safe, resilient and sustainable''~\cite{united_nations_transforming_2015}. 

Despite its significance in urban planning theories, the association between high-traffic roads and reduced social connectivity has never been measured empirically, with the notable exception of a few small-scale, survey-based studies~\cite{kemp_social_1965,appleyard1972environmental}. Previous quantitative research in this area, constrained by the scarcity of geo-referenced social network data~\cite{viry_role_2022,ye_spatial_2021}, has focused instead on measuring socio-economic segregation in cities. This goal has been achieved either by using static demographic data~\cite{jargowsky1996take,roberto_spatial_2021} or, more recently, through mobility data~\cite{xu_quantifying_2019,moro_mobility_2021,fan_diversity_2023}, with only sporadic attempts to link segregation to urban barriers~\cite{toth_inequality_2021,pinter_neighborhoods_2023}. While highly valuable, such previous research could not explicitly consider social ties. However, providing an \emph{explicit, quantitative} estimation of the barrier effect of different roads in curbing social ties is crucial for guiding evidence-based plans of restorative urban interventions and for prioritizing them according to their estimated benefits~\cite{williams_racial_2020}. 

To fill this gap, we introduce a method to systematically quantify the association between highways and social ties at multiple scales, ranging from individual highway segments to entire metropolitan areas. We focus on the network of urban highways in the US. This highway network offers a compelling subject for the study of barrier effects: with a cost of at least 1.4 trillion USD~\cite{nall_road_2018}, US highways were built to bridge city centers and newly created suburbs; simultaneously, they displaced an estimated 1 million people from their neighborhoods and today pose hard-to-cross physical barriers to pedestrians and cyclists~\cite{anciaes2016urban, miner_car_2024}.

Onto this network of urban highways within the 50 largest metropolitan areas in the US, we overlay a massive geolocated social network of ties between individuals who follow each other on Twitter~\cite{dobos_multi-terabyte_2013}. We compute a \emph{Barrier Score} which quantifies the reduction in the number of social ties crossing highways, comparing the empirical crossings with a null model that makes ties oblivious to highways. The distribution of Barrier Scores reveals that in all 50 cities, the presence of highways consistently correlates with reduced social connectivity compared to the null model, showing that urban highways are barriers to social ties. This reduction is stronger between people living closer to each other, peaking at distances below $5\,\mathrm{km}$ in most cities and fading beyond $20\,\mathrm{km}$.

Notoriously, urban highways in the US have been instrumentalized for government-backed racial segregation, creating social divides between communities that persist to this day~\cite{rothstein_color_2017, kimble_city_2024}. We therefore revisit several highways in US cities that are well-documented for their historic role in racial segregation, finding potential evidence for long-lasting effects several decades after their construction, by measuring high Barrier Scores in \emph{contemporary} social networks.

\section*{Results}\label{sec:results}

\begin{figure}
   \begin{center}
   \includegraphics[width=0.92\linewidth]{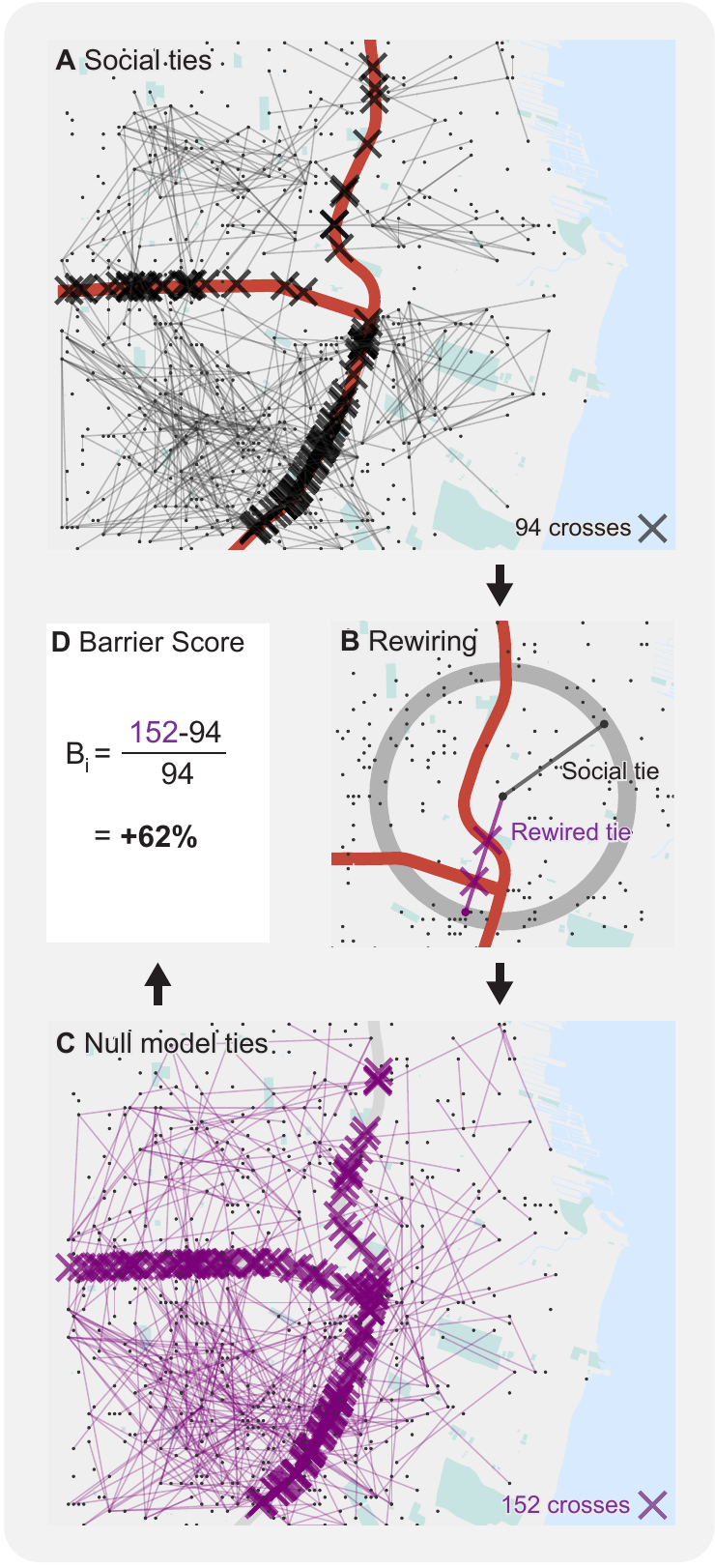}
  \caption{\textbf{The Highway Barrier Score measures the association between highways and social ties crossing them.} Calculating the Barrier Score ${B_i}$ of a highway section $i$ follows four steps. The illustrated highway section consists of highway I-94 and the \mbox{8 Mile} Road in Detroit. (\emph{A}) Social ties: Count the number of times $c_i = 94$ that social ties (grey) between home locations of individuals (grey dots) cross the highway $i$ (red). (\emph{B}) Rewiring: A spatial null model randomly rewires all social ties within a distance ring with a radius equal to the length of the original social tie. Within the ring, a random node is selected for rewiring with probability proportional to the local user population density, to reflect the spatial gravity law. The rewired null model ties remove any relationship between ties and highways because the rewiring is agnostic to highways. (\emph{C}) Null model ties: Count the number of crosses $c_i^{\mathrm{null}} = 152$ of null model ties with the highway. (\emph{D}) Highway Barrier Score: Calculate the Highway Barrier Score as ${B_i} = \frac{c_i^{\mathrm{null}}-c_i}{c_i}$. In this example, ${B_i} = +62\%$, which is the relative increase of social ties crossing the highway if ties were formed disregarding its presence. For illustration purposes, in this figure we only plot links that are fully within the view area. 
  \label{fig01:cover}
  }
  \end{center}
\end{figure}

Our starting point is a large collection of Twitter user activity from 2012-2013~\cite{dobos_multi-terabyte_2013} that contains the approximate home locations of almost 1M~Twitter users living within the boundaries of the 50 most populous metropolitan areas in the US. These users are connected by more than 2.7M~social ties representing mutual followership~\cite{bokanyi_universal_2021}. Fig.~\ref{fig:num_nodes_edges} and Table~\ref{tab:metro_information} provide detailed statistics on the data. To this social tie data we relate urban highway networks extracted from OpenStreetMap (OSM). See details in Materials and Methods.

Figure~\ref{fig01:cover}\emph{A} shows a small data sample to illustrate how we relate social ties to highway data. In this example of a particular highway section $i$, we count social ties crossing it $c_i=94$ times. Ideally, quantifying the correlation between the presence of a highway and the social ties crossing it would require to compare the frequency of social ties intersecting the highway in the empirical data against the same frequency from data collected in a hypothetical counterfactual scenario without highways. To approximate this ideal setting using observational data only, we construct a null model of the social network and compare the observed network patterns to this randomized setting (Fig.~\ref{fig01:cover}\emph{B}). Our null model rewires social ties by preserving the original degree of nodes, the distance between connected users, and the tendency of creating ties with people living in densely populated areas (Fig.~\ref{fig:gravity_model_r2}), known as the spatial gravity law~\cite{expert_uncovering_2011}. This model preserves the fundamental properties of the original social network with minimal error (Fig.~\ref{fig:null_model_error}) while disrupting any correlation with highway locations, as the model is oblivious to them. Figure~\ref{fig01:cover}\emph{C} shows the rewired version of the example ties from Fig.~\ref{fig01:cover}\emph{A}. In this example, we now count these null model ties crossing the highway section $c_i^{\mathrm{null}}=152$ times.

\begin{figure}
    \centering
    \includegraphics[width=\columnwidth]{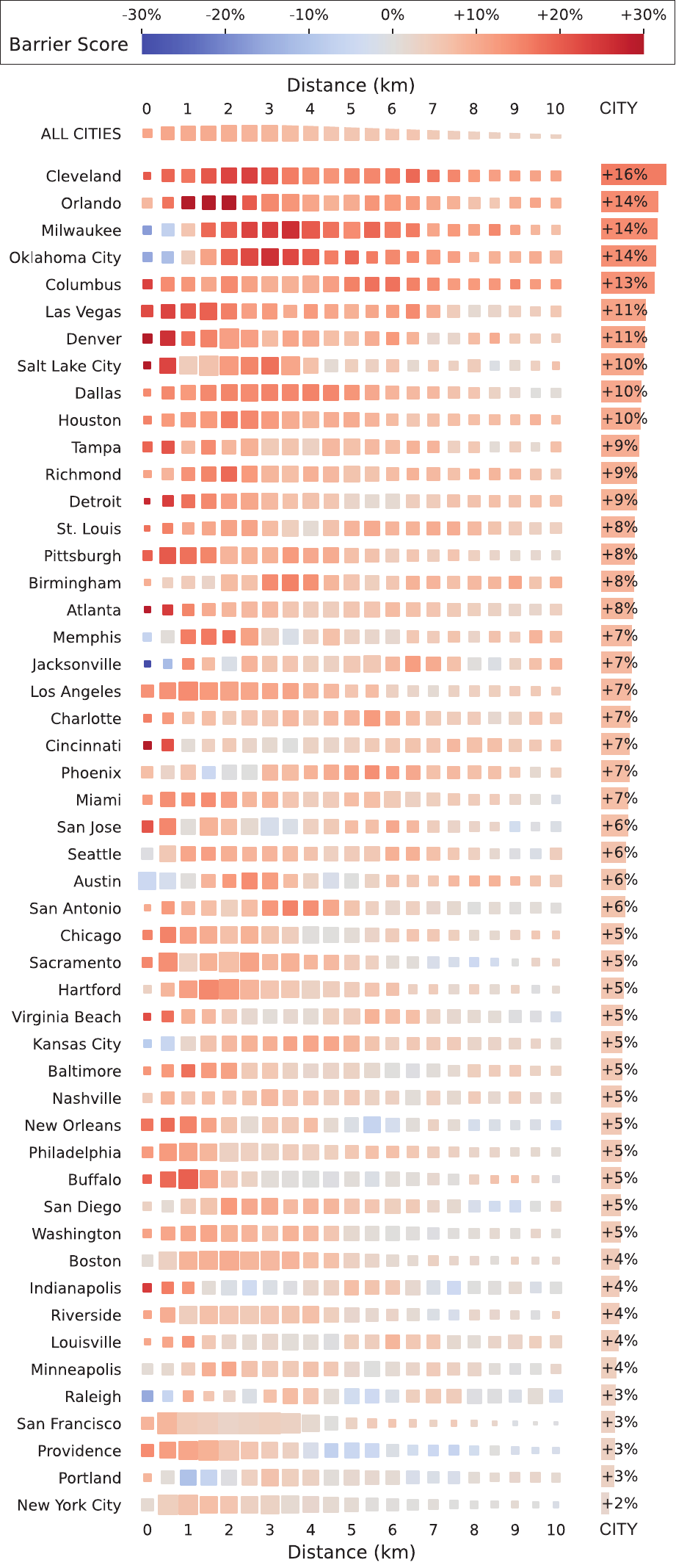}
    \caption{\textbf{The Barrier Scores across the top 50 metropolitan areas in the US are consistently positive.} \emph{(Left)} Heatmap of all Barrier Scores $B(d)$ grouped into $0.5\,\mathrm{km}$ bins of social tie distance. Color denotes Barrier Score, square size denotes the fraction of social ties in each distance band relative to all ties in the city. All cities have positive Barrier Scores over most distances. Often, there is a smoothly reached peak distance, for example in Orlando at around $d_{\mathrm{peak}} \approx 1.5\,\mathrm{km}$. The top row labelled ``ALL CITIES'' reports the distance-binned Barrier Scores averaged over all cities. (\emph{Right}) The bar plot labelled ``CITY'' reports the Barrier Score $B$ calculated considering all ties with distances up to $10\,\mathrm{km}$. All results shown are averaged over 15 randomized runs of the null model.
    \label{fig02:heatmap}}
\end{figure}

Using this null model, we define the \emph{Highway Barrier Score} ${B}_i = \frac{c_i^{\mathrm{null}}-c_i}{c_i}$ for a highway section~$i$ as the relative difference in the number of social ties crossing the section in the null model ($c_i^{\mathrm{null}}$) versus the empirical data ($c_i$). This score reflects the hypothetical increase in social ties crossing the path of the highway in its absence. Positive scores indicate that highways are associated with reduced social connectivity across the regions they bisect. In our example (Fig.~\ref{fig01:cover}\emph{D}), the Highway Barrier Score of ${B_i} = \frac{152-94}{94} = +62\%$ means that in a world where social connections are independent of the presence of highways, there are 62\% more social ties crossing the highway section~$i$.

Generalizing the Highway Barrier Score $B_i$ to a whole city, we define the \emph{Barrier Score} $B$ which aggregates the local signals across all highways and social ties over the entire metropolitan area, measuring the average increase in highway crossings per social tie in the null model relative to the observed data. This aggregate score captures a wide range of social tie distances up to $10\,\mathrm{km}$ and normalizes them appropriately; see Eq.~\ref{eqn:barrier_score} in Materials and Methods.

\subsection*{Barrier Scores are positive and diminish with distance}
The Barrier Scores $B$ for 50 cities, reported in Fig.~\ref{fig02:heatmap}~\emph{Right}, consistently show positive values, ranging from +2\% in New York City to +16\% in Cleveland, indicating that in general, highways are associated with fewer social connections in all considered cities. 

Starting from this overall city-wide score $B$, let us zoom back in, still considering all of a city's highways but limiting ourselves to social ties connecting users at a fixed distance of $d\,\mathrm{km}$. This \emph{distance-binned Barrier Score} $B(d)$ allows us to explore how the association between highway presence and reduced social connectivity varies with the geographical distance between users. The heatmap in Fig.~\ref{fig02:heatmap}~\emph{Left} shows Barrier Scores $B(d)$ calculated for social ties of fixed distance. Generally, Barrier Scores are positive (red) across most distances. They tend to peak at a relatively short distance $d_\mathrm{peak}$, for example, $d_\mathrm{peak} \approx 1.5\,\mathrm{km}$ in Orlando and $d_\mathrm{peak} \approx 3.5\,\mathrm{km}$ in Milwaukee.
At greater distances, Barrier Scores gradually diminish and at times become slightly negative (blue), meaning some highways are associated with a higher probability of social ties connecting people who live far away from each other, compared to the null model. Only occasionally, we find negative Barrier Scores at very short distances.

\begin{figure}[t!]
   \begin{center}
   \includegraphics[width=0.9\linewidth]{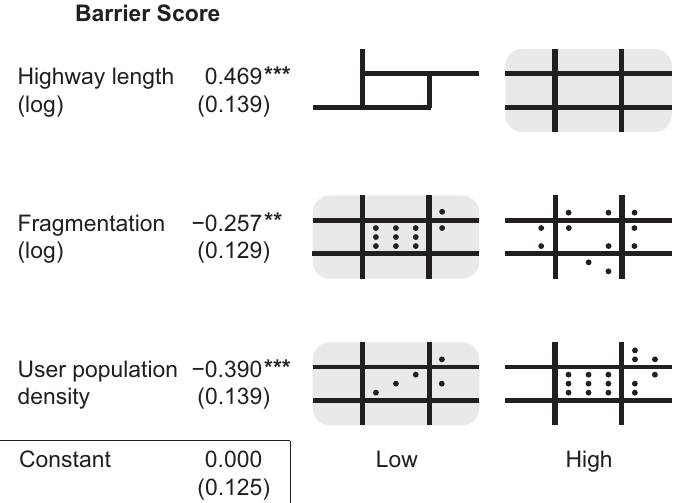}
  \caption{\textbf{Ordinary least squares regression across 50 cities reveals correlations between the Barrier Score and spatial features.} The Barrier Score increases 1)~with increasing highway length, 2)~with decreasing fragmentation, 3)~with decreasing user population density. The sketches on the right illustrate low and high values for the three features that are highway length, fragmentation, and user population density. Highways and user population are depicted via lines and dots, respectively. Grey backgrounds illustrate the signs of the regression coefficients. ***: p$<$0.01, **: p$<$0.05, *: p$<$0.1. Observations: 50. $R^2_{\mathrm{adj}}=0.231$.
  \label{fig03:regression}
  }
  \end{center}
\end{figure}

\subsection*{Regression models substantiate the barrier effect amid other factors}
To explain the city-level variation in Barrier Scores, we create a parsimonious ordinary least squares model across the 50 cities with three key explanatory variables, illustrated in Fig.~\ref{fig03:regression}: 1)~the total highway length within the metropolitan area, 2)~how much the Twitter user population is fragmented by highways, as measured by the Highway Fragmentation Index (Eq.~\ref{eqn:HFI} in Materials and Methods), and 3)~the user population density in the metropolitan area as a control variable and normalizing factor for highway length. We check the model for robustness in Fig.~\ref{fig:regression_city_sensitivity}.

The significant regression coefficients (Fig.~\ref{fig03:regression}) reveal that cities with high Barrier Scores typically have longer highway networks ($\beta=0.469$), a user population less fragmented by highways ($\beta=-0.257$), and a lower user population density ($\beta=-0.390$). These results are intuitively explained by varying each factor individually while holding the others constant. First, at same fragmentation and density of the user population, cities with a longer highway network require more frequent highway crossings to maintain social connections. Yet, the number crossings increases more rapidly for the null model than for the real social network, thus yielding higher Barrier Scores. Second, the negative coefficient of the fragmentation variable is consistent with the semantics of our null model: cities where the user population is concentrated in a few areas attract social interactions from many peripheral areas~\cite{halbert_examining_2008}, as reflected by the spatial gravity law in the null model. When these highly populated areas are separated by highways from the rest of the city, the behavior of the null model is unaffected, but the likelihood of creating a social tie that crosses a highway is comparatively lower in the empirical data, thus yielding a higher Barrier Score. Third, given the same highway length and spatial fragmentation of people, individuals have fewer opportunities to form social ties close-by~\cite{schlapfer_scaling_2014}. The resulting longer ties end up crossing more highways in the null model than in the empirical data.

\begin{table*}[th!]
\begin{center}
\begin{tabular}{@{\extracolsep{5pt}}lccccc} 
\hline \\[-1.8ex] 
\\[-1.8ex] & \multicolumn{5}{c}{\textbf{Number of social ties (log)}} \\ 
\\[-1.8ex] & (1) & (2) & (3) & (4) & (5)\\ 
\hline \\[-1.8ex] 
 Nr.~of highways crossed (log) &  & $-$0.025$^{***}$ &  &  & $-$0.021$^{***}$ \\ 
  &  & (0.001) &  &  & (0.001) \\ 
 Income abs.~difference &  &  & $-$0.019$^{***}$ &  & $-$0.018$^{***}$ \\ 
  &  &  & (0.000) &  & (0.000) \\ 
 Racial similarity &  &  &  & 0.029$^{***}$ & 0.027$^{***}$ \\ 
  &  &  &  & (0.000) & (0.000) \\ 
 Distance (log) & $-$0.101$^{***}$ & $-$0.085$^{***}$ & $-$0.099$^{***}$ & $-$0.101$^{***}$ & $-$0.086$^{***}$ \\ 
  & (0.000) & (0.001) & (0.000) & (0.000) & (0.001) \\ 
 User population (product log) & 0.029$^{***}$ & 0.027$^{***}$ & 0.026$^{***}$ & 0.026$^{***}$ & 0.022$^{***}$ \\ 
  & (0.000) & (0.000) & (0.000) & (0.000) & (0.000) \\ 
 Constant & 0.207$^{***}$ & 0.209$^{***}$ & 0.226$^{***}$ & 0.195$^{***}$ & 0.216$^{***}$ \\ 
  & (0.001) & (0.001) & (0.001) & (0.001) & (0.001) \\ 
\hline \\[-1.8ex] 
Metro fixed effect & Yes & Yes & Yes & Yes & Yes \\ 
Observations & 2,668,666 & 2,668,666 & 2,668,666 & 2,668,666 & 2,668,666 \\ 
R$^{2}$ & 0.042 & 0.043 & 0.045 & 0.047 & 0.050 \\ 
\hline 
\end{tabular}
\end{center}
\caption{Ordinary least squares regression models to predict the number of social connections between pairs of census tracts from spatial and socio-demographic features. \textmd{All the models include the metropolitan area as fixed effect. Crucially, the number of social ties between two tracts decreases with the number of highways that are crossed, after controlling for distance, user population, and socio-economic differences between the tracts. \mbox{***: p$<$0.01}, \mbox{**: p$<$0.05}, \mbox{*: p$<$0.1.}}}
\label{tab01:regression_tracts}
\end{table*}

We now complement our city-level model with multivariate regression models that describe the variability of social connectivity between census tracts. These fine-grained models allow us to verify whether the relationship between highways and social ties holds at a more granular spatial scale while controlling for local socio-economic characteristics that are known to affect social connections within cities~\cite{dong_segregated_2020}. When considering all possible tracts, many tract pairs have no highways in the space between them or no social ties connecting them, which makes it impossible to define a Barrier Score for them. Therefore, instead of considering Barrier Scores, these fine-grained models predict the observed number of social ties between pairs of tracts from five variables: 1) the average number of highways crossed by social ties between tracts, 2) the difference in average household income, 3) a dummy variable indicating whether the two tracts have the same racial majority group, and two controls for distance and user population. The sample behind the models is composed of all 2,668,666 census tract pairs that are connected by at least one social tie either in the empirical data or the null model (Table~\ref{tab:si_setting_explanation_tab}). 

The results confirm the expectation that pairs of tracts with shorter distance, higher user population, and higher socio-economic similarity exhibit more social ties. Even after adjusting for these factors (last model (5) in Table~\ref{tab01:regression_tracts}), a significant negative correlation persists between the number of highways separating tracts and the quantity of social connections ($\beta=-0.021$). Notably, the effect size of highways is comparable to that of income variables ($\beta=-0.018$) and racial similarity ($\beta=0.027$), indicating that highways may be as influential as socio-economic factors in contributing to social fragmentation. These results replicate when fitting city-specific models (Fig.~\ref{fig:reg_coeff_for_each_metro}). In SI~Section~\ref{sec:si:alt_regressions} we explain the models and variables in greater detail and corroborate the robustness of our results by experimenting with alternative models (Tables~\ref{tab:si_regressions_tracts},~\ref{tab:si_interaction_model}).

Furthermore, when examining tract pairs across fixed distances, we observe that the coefficient for the number of highways increases with distance, becoming positive beyond $d =20\,\mathrm{km}$ (Fig.~\ref{fig:si_interplots}). This pattern is consistent with the diminishing Barrier Scores over distance (Fig.~\ref{fig02:heatmap}), and it suggests that highways represent barriers to social ties predominantly at shorter spatial scales, while they may foster connectivity at longer distances.

\subsection*{Barrier Scores are consistent with racial segregation}

\begin{figure*}[th!]
    \centering
    \includegraphics[width=\textwidth]{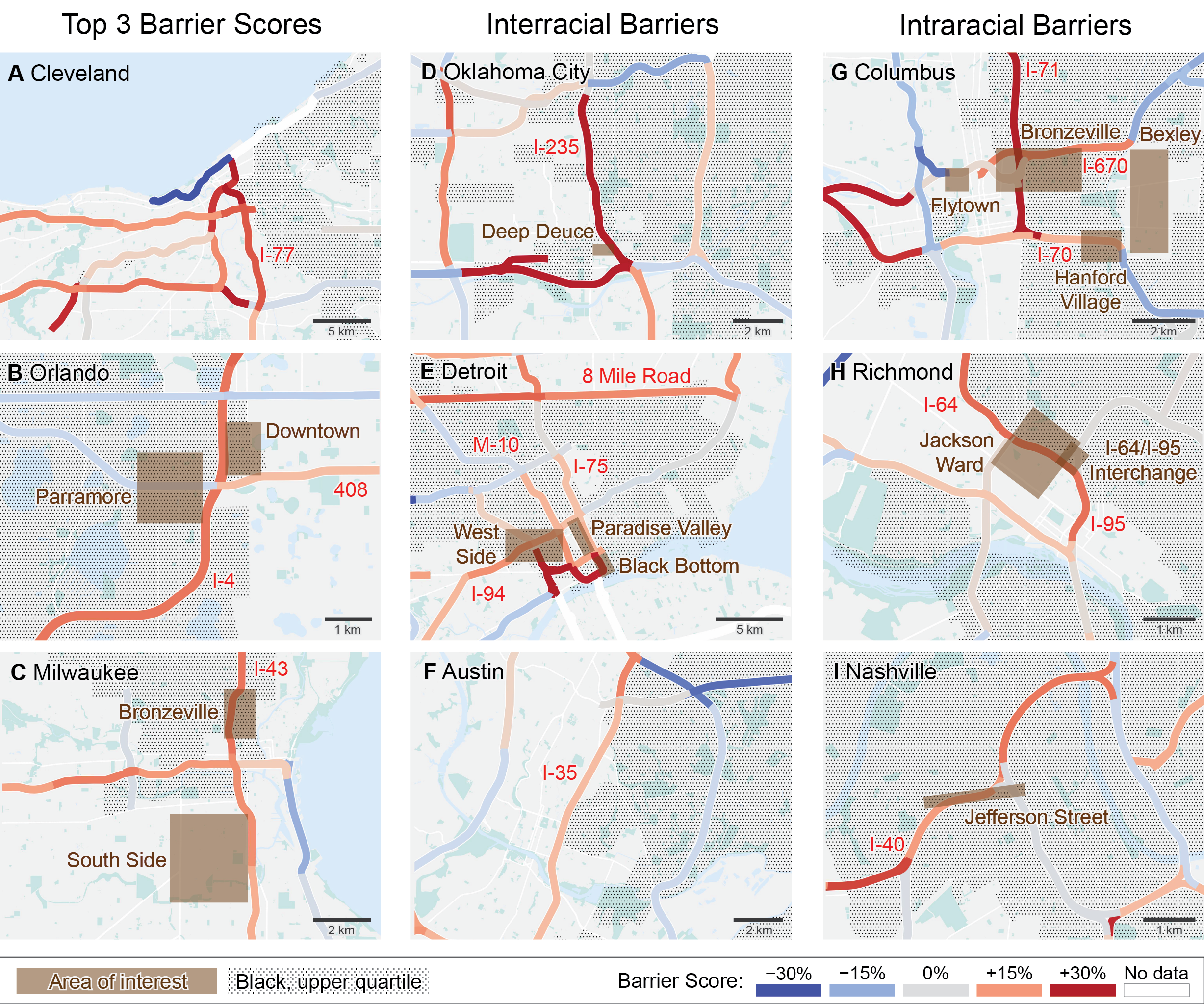}
    \caption{\textbf{Historical case studies of highways associated with racial segregation.} Highways are in color, following the color coding of Fig.~\ref{fig02:heatmap} (red: positive Barrier Score, blue: negative Barrier Score, white: insufficient data). Brown rectangles denote historically relevant areas. Black dotted areas denote a city's districts with a black population share in the upper quartile. (\emph{A,B,C}) Top 3 Barrier Scores: Cleveland, OH; Orlando, FL; Milwaukee, WI. Top Barrier Scores are consistent with these cities having well-known histories of highway-related racial segregation. (\emph{D,E,F}) Interracial Barriers: Oklahoma City, OK; Cleveland, OH; Austin, TX. The barrier between Black and non-Black neighborhoods are clearly visible around I-235, the 8 Mile Road, and I-35, respectively. Detroit additionally features intraracial barriers around M-10, I-94, and I-75. (\emph{G,H,I}) Intraracial barriers: Columbus, OH; Richmond, VA; Nashville, TN. Here the focus is on historically Black neighborhoods like Hanford Village, Jackson Ward, or Jefferson Street, respectively, that have been purposefully demolished via highway construction.}
    \label{fig04:casestudies}
\end{figure*}

To highlight the practical implications of our quantitative findings, we now frame them within a broader historical context, with a particular focus on racial residential segregation. Race is only one of many social categories that can influence the formation of social connections. However, the Interstate Highway System -- which we study here in the urban context -- is highly relevant for aggravating racial segregation in US cities \cite{mohl_interstates_2002}, making the association of our Barrier Score with racial residential segregation a compelling case study. Overwhelming historical records show how urban highway construction in the name of ``urban renewal'' has been frequently used as a racist policy toolbox to purposefully disrupt or isolate Black neighborhoods \cite{trounstine_segregation_2018}, together with other de jure segregation tools like redlining and housing policy \cite{rothstein_color_2017}. Such exclusionary urban policies, put in place decades ago, have literally cemented racial divides in US cities and have therefore not lost any of their societal relevance today \cite{arcadi_concrete_2023,kimble_city_2024}. Indeed, the US Department of Transportation acknowledges the issue in its 2023 ``Reconnecting Communities Pilot Program'', an ``initiative to reconnect communities that are cut off from opportunity and burdened by past transportation infrastructure decisions'' \cite{us_department_of_transportation_press_2023}.

The decisions on where to place new highways within the urban fabric were often racially motivated, following different considerations. Highways either could embody a policy aimed at segregating Black people from the rest of the population \cite{schindler_architectural_2014}, thus forming an \emph{interracial} barrier; or highways could be purposefully built \emph{through} Black neighborhoods, both with the intention to disrupt them and to avoid disturbances for white neighborhoods \cite{mohl_urban_2001}, thus forming an \emph{intraracial} barrier. As illustrated in Fig.~\ref{fig04:casestudies}, we therefore take a closer look at three groups of cities: cities with the highest Barrier Scores (Cleveland, Orlando, Milwaukee, in Fig.~\ref{fig04:casestudies}\emph{A-C}); cities with highways known from the historical literature as interracial barriers (Oklahoma City, Detroit, Austin, in Fig.~\ref{fig04:casestudies}\emph{D-F}); and cities with highways known as intraracial barriers (Columbus, Richmond, Nashville, in Fig.~\ref{fig04:casestudies}\emph{G-I}). Strikingly, for all case studies, highways that are historically associated with racial segregation also display high Barrier Scores. For each of these nine cities, we discuss the local historical context of highway development and its relation to racial segregation in SI~Section~\ref{sec:si:racialcontext}, summarized in the following paragraphs.

All three cities with the highest Barrier Scores, i.e., Cleveland, Orlando, and Milwaukee, have an abundant history of racial segregation by means of infrastructure. Cleveland, the city with the highest Barrier Score, is one of the poorest and most racially segregated among major US cities \cite{yankey_neighborhood_2023}. Here, the northern part of I-77 separates majority Black neighborhoods in the east from the rest of the city (Fig.~\ref{fig04:casestudies}\emph{A}). Orlando (Fig.~\ref{fig04:casestudies}\emph{B}), as of today, remains highly segregated along the I-4. The construction of the I-4 and the Expressway 408 particularly disrupted the once thriving Black neighborhood of Parramore \cite{brotemarkle_crossing_2006}. Lastly, Milwaukee (Fig.~\ref{fig04:casestudies}\emph{C}) is also a highly segregated city, with majority Black neighborhoods like Bronzeville in the North and a historically ``solidly Polish'' South Side \cite{lackey_milwaukees_2013}. Here, the construction of the I-43 disrupted and displaced numerous Black communities such as Bronzeville.

Next, we discuss the three cities with highways as interracial barriers. In Oklahoma City (Fig.~\ref{fig04:casestudies}\emph{D}), the ``urban renewal'' highway construction projects had particularly dire impacts on historically Black neighborhoods such as Deep Deuce \cite{payne_newbuild_2019}. As of today, the I-235 in Oklahoma City remains a clearly perceived division line between majority Black and majority white neighborhoods \cite{felder_citys_2014}. In Detroit (Fig.~\ref{fig04:casestudies}\emph{E}), the construction of several highways during ``urban renewal'' erased and eroded numerous historically Black neighborhoods such as Black Bottom and Paradise Valley \cite{sugrue_origins_2014}. Here, ``expressway displacement'' \cite{sugrue_origins_2014} combined with pronounced discrimination led to several housing crises over the last decades, severely impacting the Black population. Lastly, in the city of Austin (Fig.~\ref{fig04:casestudies}\emph{F}), the I-35 was built along East Avenue, an intentionally enforced segregation line whose impacts are visible up to this day \cite{skop_austin_2010}. At the same time, the I-35, for which expansion plans are currently underway with 4~billion~USD allocated \cite{bernier_your_2024,kimble_city_2024}, stands out with a high Barrier Score.

Finally, three cities from our case studies are well-known for their intraracial highway barriers. Columbus (Fig.~\ref{fig04:casestudies}\emph{G}) is a particularly startling example of highway construction as deliberate neighborhood destruction \cite{thompson_how_2020}, with today's highway routes aligning with former redlining maps. The most severely impacted neighborhoods like Flytown, Hanford Village, or Bronzeville, were economically disadvantaged and predominantly Black; at the same time, the closeby but predominantly white, affluent neighborhood Bexley was spared from the highways \cite{smith_african-american_2014, thompson_how_2020}. In Richmond (Fig.~\ref{fig04:casestudies}\emph{H}), highway construction and segregationist housing policies interacted to create a ``concentration of racialized poverty'' \cite{howard_reframing_2016} that lasts until the present day. Richmond's neighborhood of Jackson Ward, formerly dubbed ``Black Wall Street'', was bisected by the I-95 and the I-64/I-95 interchange, ultimately leading to its decline. Finally, in Nashville (Fig.~\ref{fig04:casestudies}\emph{I}), the I-40 was routed through a bustling Black neighborhood without any appraisal of potential consequences for the community, bisecting the once-thriving Jefferson Street, and at a larger scale undermining Black commercial and educational institutions, decisively contributing to today’s high poverty rates in the area \cite{haynes_one_2020}.

This historic contextualization is highly relevant in connection with our research. We find for all these nine cities that historic spatial divides are reflected in our contemporary analysis of social ties: all investigated highways display high Barrier Scores. While a broader, systematic investigation that checks every possible highway section and historical note is outside of the scope of our research, these findings add another piece of evidence consistent with the established concept that urban highways in the US have a strong relation with government-backed racial segregation~\cite{rothstein_color_2017}. Now our research additionally shows that reduced social connectivity in the presence of highways \emph{can be quantitatively detected at high resolution}. 

\section*{Discussion and conclusion}\label{sec:discussion}

To gauge the robustness of our results, we conduct three experiments. First, to check that high Barrier Scores are specific to highways among all street types, we replicate the analysis on other categories of roadways. While these road types also yield positive Barrier Scores, they are markedly lower than those associated with highways (Fig.~\ref{fig:streettypes}). For example, for the lowest distance $d=0.5\,\mathrm{km}$, $B(d)$ for highways is around +12\%, while it is +8\% for primary roads, +5\% for secondary roads, and +4\% for residential streets. The $B(d)$ values decrease with distance and retain this order. 
Lower Barrier Scores for less trafficked streets are intuitive, as such streets can be easier crossed on foot, corroborating urban planning literature which suggests that the traversability of streets influences social connectivity~\cite{appleyard_livable_1981,grannis_importance_1998}.

Second, we check whether the higher Barrier Scores for highways might be due to their lower total length compared to other street types. To control for this length imbalance, we recalculate the Barrier Scores using a simulated, randomized version of the highway network that preserves the total length of highways but alters their spatial distribution (see SI~Section~\ref{sec:si:nullmodel_highways}). The comparison between empirical and randomized highway layouts reveals significantly reduced Barrier Scores in the randomized scenarios (Fig.~\ref{fig:random_street_layout}), confirming that the spatial positioning of highways plays a more important role than their total length.

Third, we replicate our findings on a distinct social network: Gowalla. It is a location-based social network platform where users connect with friends and share their own location with them through check-ins~\cite{cho_friendship_2011}. The Gowalla dataset contains five cities with sufficient data coverage (Table~\ref{tab:si_gowalla_stats}). The Barrier Scores derived from Gowalla ties are notably higher than those from Twitter across all distances (Fig.~\ref{fig:gowalla}). Considering Gowalla's emphasis on fostering real-life interactions among users (its mission being ``keep up with your friends in the real world.''), it is reasonable to infer that this platform's social ties might be inherently stronger than the ties on Twitter which does not have this emphasis. This observation suggests that the interplay between highways and social connection may be even more pronounced for stronger social ties.

Being first of its kind, our work does not cover additional aspects of the relationship between social connectivity and spatial features open for future research. The relationship between highways and social connectivity is potentially subject to confounding factors such as social dynamics, terrain morphology, or public transit~\cite{roberto_spatial_2021,toth_inequality_2021,pappalardo_future_2023}. The Barrier Score we derived likely reflects a composite influence of these elements, and more refined spatial null models could help to disentangle them. Furthermore, our null model provides a somewhat reductive perspective on the interplay between social networks and highways. For example, it does not distinguish cases where a highway walls off two individuals from cases where it facilitates them to connect. Additionally, the study's observational design means that our null model is limited to considering rewiring of \emph{existing} social ties, so it cannot account for the possibility of ties appearing or vanishing in the absence of highways. Lastly, our reliance on social media data limits representativeness~\cite{Mislove2011}, a well-documented issue in social media research~\cite{Sloan2015}. Although we found a strong correlation between user volume and user population size across the 50 cities studied, and our data covers a set of tracts that is representative of the distribution of income (Fig.~\ref{fig:representativeness}), our findings may not be generalizable to the entire population of these areas.

In conclusion, by going beyond demographic approaches, observing social ties \emph{explicitly}, we have shown that there is a quantifiable association between urban highways and reduced levels of social connectivity, especially at short distances. Our analysis adds a highly granular perspective to former work, corroborating and quantifying the intuition that urban highways are indeed barriers to social ties. At the same time our analysis also indicates that highways can facilitate connecting people at larger distances. However, this potential benefit comes with perpetuating car dependency via sprawl and induced demand~\cite{oecd_transport_2021}, and with a wide array of considerable harms~\cite{miner_car_2024} including traffic violence, environmental damage, social isolation and injustice. The social harms are corroborated by our nine historical case studies which illustrate that highway barrier effects may be considerable and long-lasting.

To be clear, our approach is so far strictly correlational and cannot establish causality: from static data it is impossible to determine how thinned-out social ties across a highway section already were before its construction, say because of an existing racial divide~\cite{sugrue_origins_2014,rothstein_color_2017}; or to which extent a new highway \emph{caused} social ties to thin out. Scrutinizing causality would require longitudinal data, for example before and after the construction or removal of an urban highway. Nevertheless, within the historical context, our results paint a clear picture. Thus, our research could already help remediate previous political failures~\cite{schindler_architectural_2014,rothstein_color_2017} and enrich the debate on contemporary highway policies~\cite{bernier_your_2024,us_department_of_transportation_press_2023,kimble_city_2024}, to account for exclusionary effects of infrastructure, and to inform reparative justice approaches~\cite{archer_white_2020, williams_racial_2020}. More generally, our research contributes to a more careful, evidence-based consideration of the social fabric in urban planning.

\matmethods{

\subsection*{Social network}\label{sec:methods:twitterdata}

We rely on an existing collection of geo-referenced tweets posted between 2012 and 2013, when the Twitter mobile app's default setting was to annotate all tweets with the precise geographic coordinates at the time of posting. Previous work~\cite{dobos_multi-terabyte_2013} used the friend-of-friend algorithm to identify the home locations of users with a sufficient number of posts with high accuracy. The dataset comes with the full network of mutual followership among all users whose home location is within the 50 most populous metropolitan areas in the United States. Overall, the network contains 982,459 users and 2,711,185 social ties between them. This dataset has proven to be a reliable resource to study spatial social networks within cities~\cite{bokanyi_universal_2021, kovacs_income-related_2022}. The home location estimation procedure, present statistics on the data, and its representativeness are described in detail in SI~Sections~\ref{sec:si:twitter_data} and \ref{sec:si:homelocationestimation}.

From the spatial perspective, we model social ties as straight segments connecting the home locations of two users. We considered the shortest path between home locations as an alternative spatial representation and found very similar results, as the length of the straight segments strongly correlates with walking distance in all cities ($\rho > 0.95$, see Fig.~\ref{fig:corr_beeline_walking}).

\subsection*{Street network}\label{sec:methods:highways}

We obtain the street network data for all 50 metropolitan areas of this study from the open and crowd-sourced platform OpenStreetMap (OSM)~\cite{openstreetmap_contributors_openstreetmap_2023}. We refer to the \textit{highway network} as the network of highways (freeways, motorways, interstates), and obtain the corresponding data from OSM by filtering street network segments by their \texttt{highway} tag attribute. The street network geometries are further simplified with OSMnx, and for the case studies, manually in QGIS (see SI~Section~\ref{sec:si:street_data} for details on OSM queries and simplification). To determine the number of social ties crossing highways, we perform a spatial join between the social ties and the highway network, and obtain the intersection points.

\subsection*{Spatial null model}\label{sec:methods:nullmodel}

Our null model is based on the \emph{Directed Configuration Model} (DCM)~\cite{newman_random_2001}, a widely-used graph randomization method that re-wires links at random while preserving the nodes' degree. To also preserve the spatial patterns of connectivity, we augment the DCM with the spatial \emph{gravity model}, an empirical relationship stating that the volume of social connections between two areas is proportional to the number of inhabitants, and inversely proportional to their distance~\cite{scellato_socio-spatial_2011}. In practice, we follow an iterative procedure in which each tie $(i,j)$ is rewired to form a new tie $(i,k)$ such as user $k$ is 1) approximately at the same distance from $i$ as $j$ is ($d_{ij} = d_{ik}$), and 2) it is selected among all candidate nodes with probability that is proportional to the density of other users around it. Details on the algorithm and its properties are discussed in the Supplementary Information.

Overall, the algorithm generates a random social network that retains both  spatial and social connectivity patterns of the original data, while disregarding any spatial elements between the two endpoints of a social connection.

\subsection*{Barrier Score}\label{sec:methods:barrier_score}
 
Consider a set $E$ of social ties $(i,j)$, each characterized by the Euclidean distance $d_{ij}$ between user $i$ and user $j$. We denote with $c_{ij}$ the number of highways that a tie $(i,j)$ crosses. We count the average number of highways that ties in $E$ cross by unit distance:
\begin{equation}
c_{E} = \frac{1}{|E|} \sum_{(i,j) \in E}\frac{c_{ij}}{d_{ij}}.
\label{eqn:avg_crossing}
\end{equation}
Intuitively, to calculate the Barrier Score, one could directly contrast the number of crosses in the real social network $c_{E}$ with the same number calculated in the randomized null model $c^{\mathrm{null}}_{E}$. In practice, the relationship between $c_{E}$ and $c^{\mathrm{null}}_{E}$ varies considerably when considering social ties across different ranges of length, and tends to converge to 0 when all long-range social ties are considered (as hinted at by Fig.~\ref{fig02:heatmap}). Therefore, to characterize cities with a score that represents all distances equally, we first compute a distance-binned Barrier Score for ties connecting users whose distance is within a distance bin $d$:
\begin{equation}
B(d) = \frac{c_E^{\mathrm{null}}(d) - c_E(d)}{c_E(d)},
\label{eqn:barrier_score_d}
\end{equation}
and then compute a final Barrier Score as an average over all $k$ distance bins up to a maximum distance $D$:
\begin{equation}
B_{\leq}(D) = \frac{1}{k} \sum_{d=0}^{D} B(d).
\label{eqn:barrier_score}
\end{equation}
We set the width of distance bins to $0.5\,\mathrm{km}$; therefore, for example, $B(2)$ considers all social ties of length between $2\,\mathrm{km}$ and $2.5\,\mathrm{km}$. To define the city-wide Barrier Score in the main results we use $10\,\mathrm{km}$ as the reference value of $D$ and refer to it simply as $B \coloneqq B_{\leq}(10)$. A sensitivity analysis of the results of regression models across different values of $D$ is reported in SI~Section~\ref{sec:si:dist_barrier_interplots}.

\subsection*{Spatial fragmentation} \label{sec:methods:spatial_fragmentation}

We measure the spatial fragmentation of a metropolitan area by highways using a modified version of the Railroad Division Index (RDI)~\cite{ananat_wrong_2011}:
%
\begin{equation}
    RDI = 1 - \sum_{i} \left(\frac{\mathrm{area}_{i}}{\mathrm{area}_{\mathrm{total}}}\right)^{2}
\end{equation}
\noindent where $\mathrm{area}_{i}$ is the area of the $i$-th subunit of fragmented space, enclosed by highways. In line with the RDI definition, we derive the subunits within a city by first combining the highway network and the metropolitan urban area boundaries and then polygonizing their spatial union~\cite{fleischmann_shape-based_2024}. To account for user population density, we weight areas by the number of users living in them, and define the Highway Fragmentation Index as: 
\begin{equation}
    H\!F\!I = 1 - \sum_{i} \left(\frac{\mathrm{users}_{i}}{\mathrm{users}_{\mathrm{total}}}\right)^{2}
    \label{eqn:HFI}
\end{equation}
\noindent A minimum fragmentation index of 0 describes a city where all residents could reach each other without crossing any highway, whereas a maximum fragmentation close to 1 denotes a city where the user population is spread uniformly across areas that are enclosed by highways.

}

\showmatmethods{}

\acknow{We thank Roberta Sinatra, Trivik Verma, Frank Neffke, and Szabolcs-Endre Horvát for helpful comments. L.M.A.~acknowledges funding from Carlsberg Foundation Project COCOONS (Grant ID: CF21-0432). M.S.~acknowledges funding from EU Horizon Project JUST STREETS (Grant agreement ID: 101104240). S.J.~acknowledges funding from EU Marie Skłodowska-Curie Postdoctoral Fellowship Programme (Grant number 101062606).}

\showacknow{}

\clearpage


\noindent{\huge \textbf{Supplementary Information}}

\setcounter{figure}{0}
\setcounter{table}{0}
\setcounter{equation}{0}
\renewcommand{\thefigure}{SI\arabic{figure}}
\renewcommand{\thetable}{SI\arabic{table}}
\renewcommand{\theequation}{SI\arabic{equation}}

\fontsize{9}{11}\selectfont

\subsection{Social network data}\label{sec:si:twitter_data}

\begin{figure}[t!]
   \centering
      \includegraphics[width=0.99\linewidth]{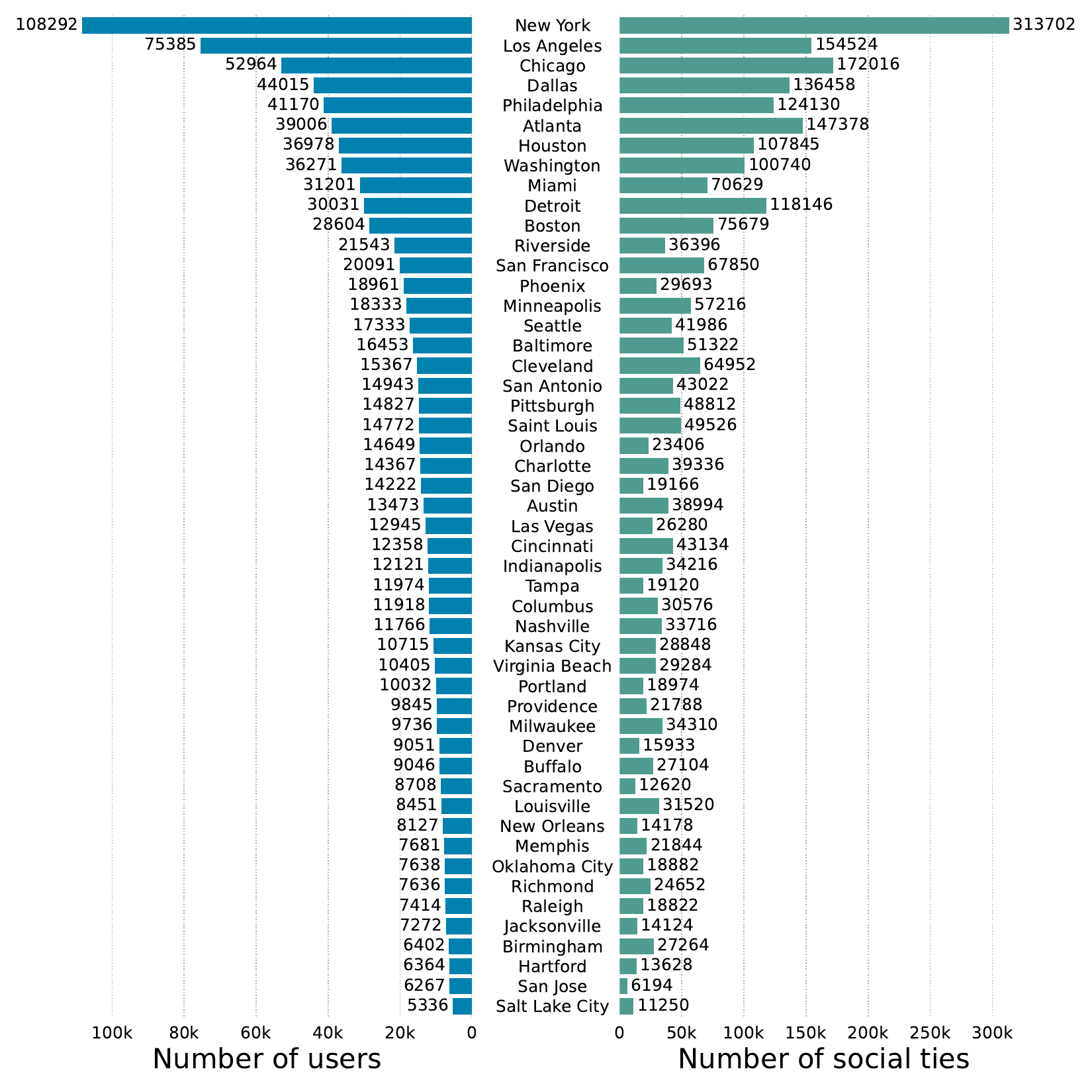}
      \caption{\textbf{Social network size.} Number of nodes and edges in the Twitter social network for the 50 largest metropolitan areas of the US.}
   \label{fig:num_nodes_edges}
\end{figure}

Social connections within cities are mapped through a large social network snapshot from Twitter with precise geolocation information. This dataset provides a remarkable context for studying spatial network patterns inside cities, as individual-level social connections are rarely available with precise geographic location at large scales. Fig.~\ref{fig:num_nodes_edges} shows the network size in terms of nodes and edges in all 50 metropolitan areas under study. Edges are based on mutual followership. Besides presenting the raw numbers of social network nodes and edges for each metropolitan area, Table~\ref{tab:metro_information} reports the bounding box coordinates we used for our modeling.

\begin{figure}[t!]
   \centering
      \includegraphics[width=0.99\linewidth]{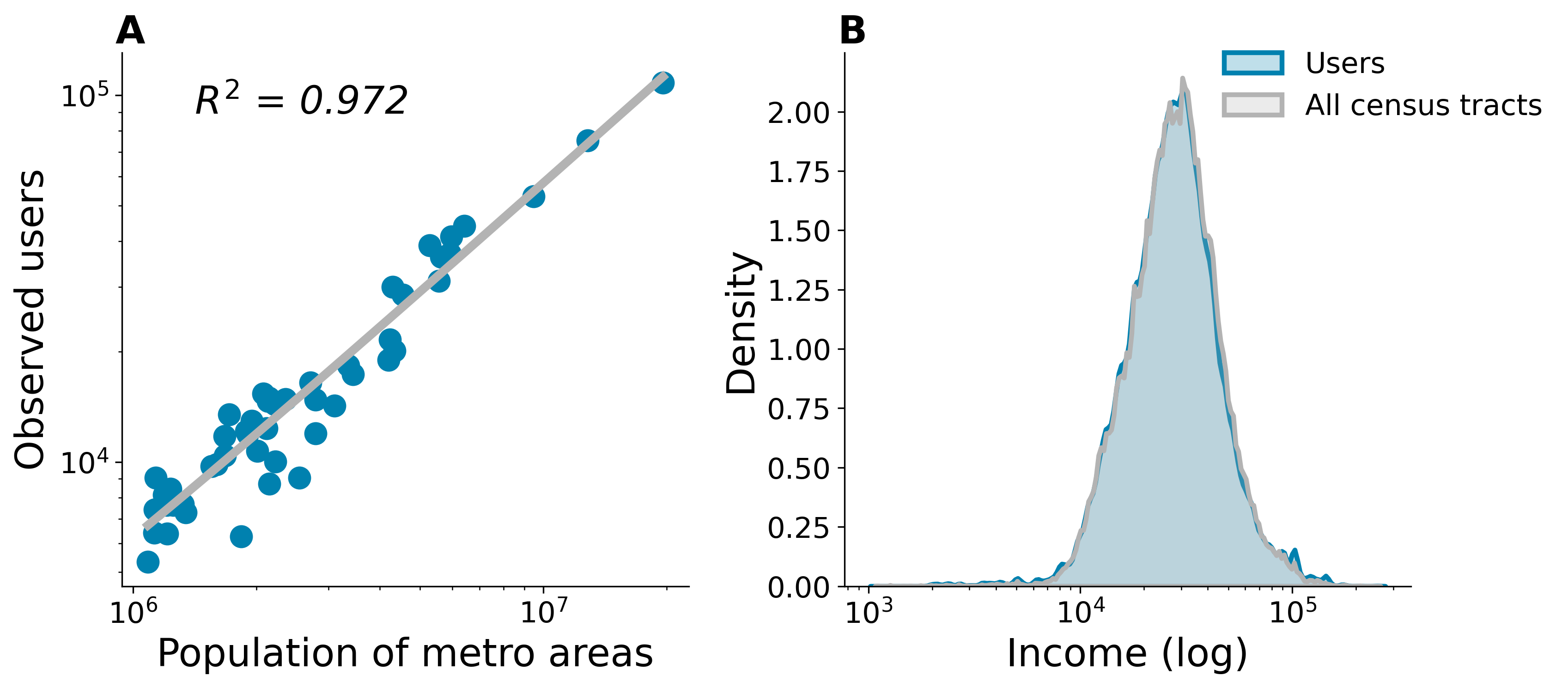}
      \caption{\textbf{Representativeness of the Twitter data.} \textbf{A.} Correlation between the population and the number of users we observe in each of the 50 largest metropolitan areas of the US. \textbf{B.} Income distribution at the home location of our observed users and at all the census tracts in the top 50 metropolitan areas. All census tracts refers to the population-weighted observation of census tract level income.}
   \label{fig:representativeness}
\end{figure}

Fig.~\ref{fig:representativeness}A illustrates that the distribution of users across cities follows closely the distribution of population size from the census. In addition, we infer the socioeconomic status of users by linking their home location to census tract-level data on income, education, and racial composition from the 2012 American Community Survey. We estimate the distribution of income of our Twitter user sample by assigning to each user the average income of the census tract corresponding to their home location. We then compare the user income distribution with the income distribution across all census tracts in those metropolitan areas weighted by population, and find them highly overlapping (Fig.~\ref{fig:representativeness}B). These checks demonstrate consistency between the total population and our sample of Twitter users.

\subsection{Home Location Estimation}\label{sec:si:homelocationestimation}

We start with the friend-of-friend algorithm~\cite{si_dobos_multi-terabyte_2013} to identify users' home locations from geocoded tweets, at a precision below census tract level on grid cells of approximately \mbox{$400\times400\,\mathrm{m}$}. This algorithm starts by identifying the three densest spatial clusters of geocoded datapoints for each user. Two geotagged tweets of the same user are considered to belong to the same spatial cluster if they are at most 1 km apart. To eliminate outliers, we iteratively filter out from each cluster the datapoints that are most distant from the cluster centroid, until all points are at most within a 3$\sigma$ radius from the centroid. After this trimming process, the three highest cardinality clusters per user are retained. 

We only keep users who have at least two of their three clusters falling within the same US metropolitan area which both contain at least 50 tweets posted during weekdays across the 2-year period covered by the dataset. In line with established practices~\cite{si_mcneill_estimating_2017, si_lambiotte_geographical_2008, si_bokanyi_universal_2021}, we label the cluster with the most tweets between 8PM and 8AM as the user's home location.

\subsection{Street network data}\label{sec:si:street_data}

We download the street network data from OpenStreetMap (OSM) \cite{si_openstreetmap_contributors_openstreetmap_2023} using the OSMnx Python package v.1.2.1~\cite{si_boeing_osmnx_2017}. Specifically, we use the \texttt{osmnx.graph.graph\_from\_bbox} function with the parameters \texttt{network\_type = all\_private}, \texttt{simplify = False}, \texttt{retain\_all = True}, \texttt{truncate\_by\_edge = True}, \texttt{clean\_periphery = False}. For each city, we construct the network comprising all the streets within the bounding box of their Metropolitan Statistical Area (MSA) as defined by the US Census Bureau (Table~\ref{tab:metro_information}). 

All OSM street segments are labeled with the OSM attribute \texttt{highway}, which identifies the type of street that the segment represents. For each city, we extract the network of highways, freeways and major transportation roads such as interstates, by considering segments labeled as \texttt{highway=motorway}, \texttt{highway=trunk}, \texttt{highway=motorway\_link}, or \texttt{highway=trunk\_link}. 

We simplify the resulting graphs with the OSMnx function \texttt{simplification.simplify\_graph} to remove interstitial nodes. The resulting street network for each city is a graph with edges representing street segments and nodes representing the intersections between them. For the qualitative studies of the 9 cities presented in the case study (Fig.~4 in the main text), we use the QGIS software~\cite{si_qgisorg_qgis_2023} to additionally simplify the graphs, manually removing truncated street segments and merging small consecutive highway segments between intersections into larger segments.

\subsection{Spatial null model detailed description}\label{sec:si:null_model}

\begin{figure}[t!]
   \centering
   \includegraphics[width=0.99\linewidth]{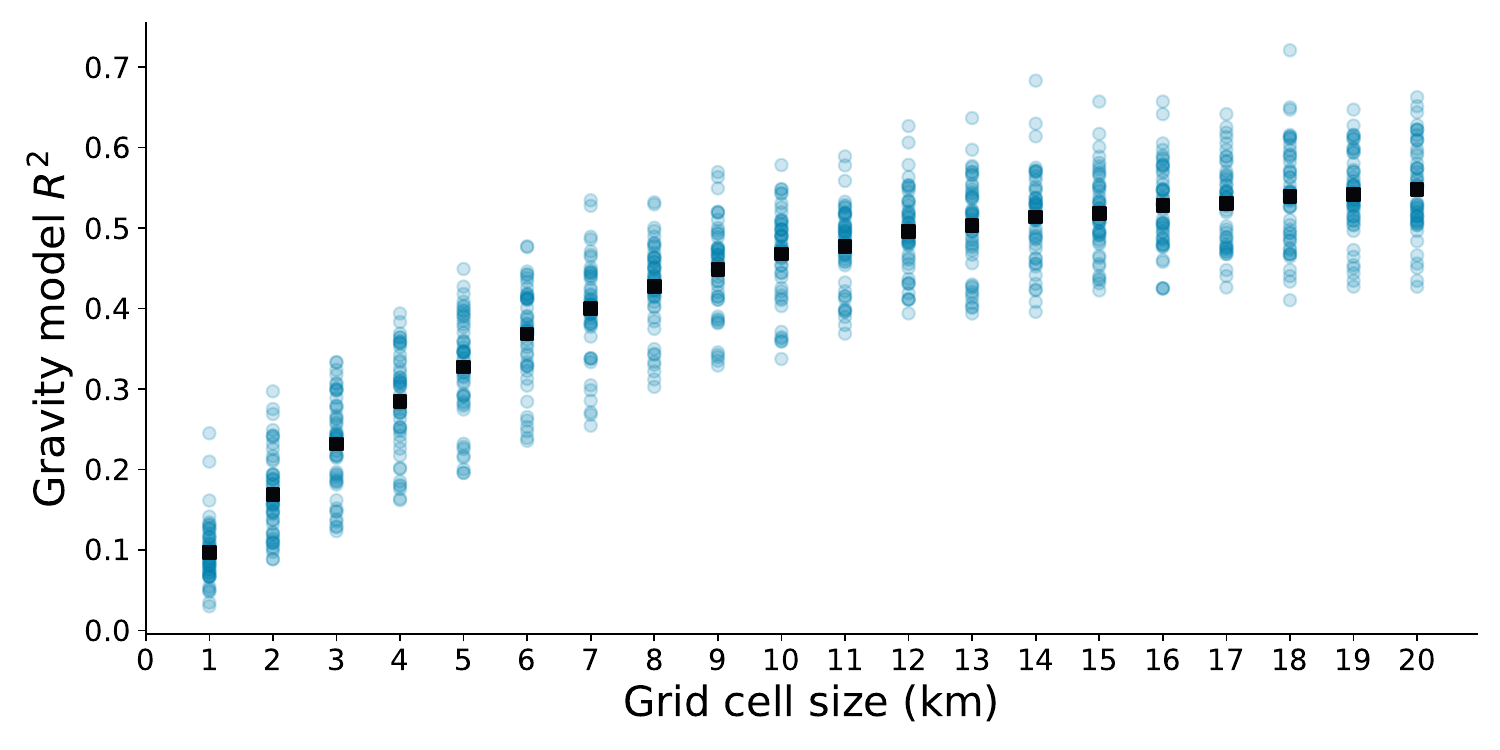}
   \caption{\textbf{Gravity model of Twitter data.} $R^2$ goodness of fit of a linear regression to estimate the number of social connections between two areas from their geographical distance and their respective number of Twitter users' home locations (Equation~\ref{eq:gravity_model_regression}). A good fit indicates that the geographical patterns of the social connections are compatible with a gravity law. Each point on the plot represents the result a regression ran on a single city and considering the tiles of a regular grid with a fixed granularity. The black squares represent the average of all the realizations for the given grid granularity. The fit stabilizes at around $R^2=0.5$ when considering tiles of $10\,\mathrm{km}$ of side.
  }
   \label{fig:gravity_model_r2}
\end{figure}

The \emph{Directed Configuration Model} (DCM)~\cite{si_newman_random_2001} is a network null model suited for directed social graphs, and has been used extensively in network science. To randomize the connections in an existing directed graph, DCM converts all incoming and outgoing edges of a node into in- and out-stubs, namely `dangling' edges attached to the node. Then, each out-stub is matched with an in-stub selected uniformly at random to form a directed edge. This method generates a new random network characterized by the same degree sequence as the original network. However, when dealing with a network in which nodes occupy a position in physical space, i.e.~a spatial network, using a standard configuration model is not sufficient, as it does not consider the spatial constraints which can heavily influence connectivity patterns. In particular, social connectivity on commonly modeled by the \emph{gravity model}~\cite{si_krings_urban_2009, si_scellato_socio-spatial_2011}, an empirical relationship inspired by the Newtonian law of gravity stating that the volume $w_{ij}$ of social connections between two geographical areas $i$ and $j$ is proportional to the total number of possible connections between them (calculated as the product between the populations in the two areas $N_i \cdot N_j$), and inversely proportional to a power of their Euclidean distance~$d_{ij}^\gamma$:
\begin{equation}\label{eq:gravity_model}
w_{ij} = \frac{N_i^{\alpha}N_j^{\beta}}{d_{ij}^{\gamma}}
 \end{equation}
The exponents of the gravity formula can be estimated from real data by fitting it to a linear regression using the Ordinary Least Squares (OLS) method: 
\begin{equation}\label{eq:gravity_model_regression}
log(w_{ij}) = \alpha \log(N_i) + \beta \log(N_j) - \gamma \log(d_{ij})
 \end{equation}
We verify empirically that the geographic arrangement of the nodes and ties in our Twitter data is compatible with the gravity model (Fig.~\ref{fig:gravity_model_r2}).

Our null model follows the algorithm of the configuration model and extends it with the gravity model, such that both degree sequences and spatial connectivity patterns are preserved. First, all ties in the network are turned into in- and out-stubs. Then, an out-stub $i$ is selected at random. Let $j$ be the stub that was originally connected to $i$, and $d_{ij}$ be the Euclidean distance between them. The set of candidate in-stubs for the random rewiring of $i$ is now restricted to those that are approximately at distance $d_{ij}$ from $i$, namely the set of stubs $S = \{k | d_{ij} - \epsilon \leq d_{ik} \leq d_{ij} + \epsilon \}$. We set empirically $\epsilon = 50m$. The matching in-stub $k$ is randomly selected among all candidates with probability that is proportional to their local density $N_k$, namely the the number of nodes in the area surrounding the node with that stub. Such an area is empirically defined as a circle of radius $500m$ centered around the candidate node. This selection based on local density reflects the gravity law in Equation~\ref{eq:gravity_model}. From that Equation, our algorithm can disregard both $N_i$, because the out-node $i$ is fixed, and $d_{ij}$ because all candidate nodes are selected to be at approximately distance $d_{ij}$ from $i$. Last, the re-wired stubs are removed from the data. The algorithm iterates over the remaining in-stubs until none are left.

As long as the set of candidate in-stubs $S$ is not empty at any iteration of the algorithm (i.e., at least one candidate in-stub is found at the desired distance), the rewired social network produced by this algorithm is a null model that exhibits strong properties; not only it reproduces the degree sequence and spatial connectivity patterns of the original network, but it also preserves the length distribution of social ties departing from \emph{each} individual node. However, this last property is not always strictly guaranteed, since the iterative nature of the algorithm may lead to the exhaustion of the set $S$ if all suitable in-stubs for the current out-stub have been used in prior iterations. When such an event occurs, the algorithm incrementally increases the buffer size $\epsilon$ until $S$ contains at least one element. In the extreme case where $\epsilon \rightarrow \infty$, all available stubs are considered with selection probability proportional to their local density, reflecting a global gravity law. 

\begin{figure}[t!]
   \centering
      \includegraphics[width=0.99\linewidth]{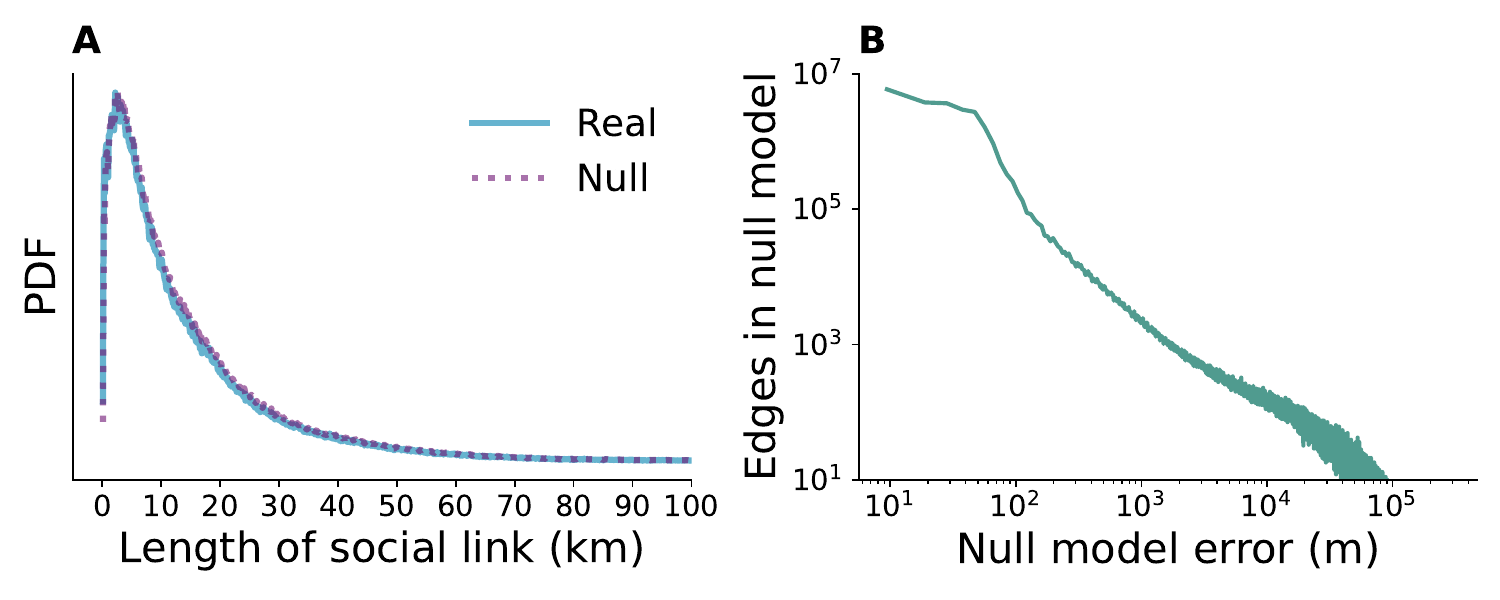}
      \caption{\textbf{Null model error.} \textbf{A.}~Probability density function of the geographical length of Twitter social ties in the real data and in the null model. The two distributions are indistinguishable according to a paired Kolmogorov-Smirnov test ($p=0.0$). \textbf{B.}~Distribution of the absolute error on the geographical length of social ties introduced by the null model. The error is calculated as the absolute difference between the length of a real tie and the length of the corresponding tie in the null model.}
   \label{fig:null_model_error}
\end{figure}

Our empirical observations indicate that the difference between the length $d_{ij}$ of an original social tie and the length $d_{ik}$ of its randomly rewired counterpart is minimal. The distributions of lengths in the real and null models are indistinguishable according to a paired Kolmogorov-Smirnov test (p = 0.0). The median error is 26m, and it is at most 100m for 92\% of the ties. In relative terms, 90\% of the null ties have a length within a 2\% error compared to their corresponding real ties. The distributions of tie lengths and null model errors are provided in Fig.~\ref{fig:null_model_error}.

\begin{table*}
\centering
\begin{tabular}{rllrr|rrrr}
\toprule
 Cbsacode & City & State &  \#Nodes &  \#Edges &     West &  South &  East &  North \\
\midrule
        12060 & Atlanta & GA & 39006 & 147378 & -85.387 & 32.845 & -83.269 & 34.618 \\ 
        12420 & Austin & TX & 13473 & 38994 & -98.298 & 29.631 & -97.024 & 30.906 \\ 
        12580 & Baltimore & MD & 16453 & 51322 & -77.312 & 38.711 & -75.748 & 39.722 \\ 
        13820 & Birmingham & AL & 6402 & 27264 & -87.422 & 32.660 & -86.044 & 34.260 \\ 
        14460 & Boston & MA & 28604 & 75679 & -71.899 & 41.566 & -70.323 & 43.573 \\ 
        15380 & Buffalo & NY & 9046 & 27104 & -79.312 & 42.438 & -78.460 & 43.635 \\ 
        16740 & Charlotte & NC & 14367 & 39336 & -81.538 & 34.458 & -79.848 & 36.059 \\ 
        16980 & Chicago & IL & 52964 & 172016 & -88.942 & 40.737 & -86.929 & 42.670 \\ 
        17140 & Cincinnati & OH & 12358 & 43134 & -85.299 & 38.473 & -83.673 & 39.729 \\ 
        17460 & Cleveland & OH & 15367 & 64952 & -82.348 & 40.988 & -81.002 & 42.252 \\ 
        18140 & Columbus & OH & 11918 & 30576 & -83.653 & 39.362 & -82.024 & 40.713 \\ 
        19100 & Dallas & TX & 44015 & 136458 & -98.067 & 32.052 & -95.859 & 33.434 \\ 
        19740 & Denver & CO & 9051 & 15933 & -106.210 & 38.693 & -103.706 & 40.044 \\ 
        19820 & Detroit & MI & 30031 & 118146 & -84.158 & 42.028 & -82.334 & 43.327 \\ 
        25540 & Hartford & CT & 6364 & 13628 & -73.030 & 41.178 & -72.099 & 42.039 \\ 
        26420 & Houston & TX & 36978 & 107845 & -96.622 & 28.765 & -94.353 & 30.630 \\ 
        26900 & Indianapolis & IN & 12121 & 34216 & -87.015 & 39.048 & -85.576 & 40.380 \\ 
        27260 & Jacksonville & FL & 7272 & 14124 & -82.460 & 29.622 & -81.151 & 30.830 \\ 
        28140 & Kansas City & MO & 10715 & 28848 & -95.188 & 38.026 & -93.477 & 39.789 \\ 
        29820 & Las Vegas & NV & 12945 & 26280 & -115.897 & 35.002 & -114.043 & 36.854 \\ 
        31080 & Los Angeles & CA & 75385 & 154524 & -118.952 & 32.750 & -117.413 & 34.823 \\ 
        31140 & Louisville & KY & 8451 & 31520 & -86.330 & 37.806 & -84.867 & 38.784 \\ 
        32820 & Memphis & TN & 7681 & 21844 & -90.589 & 34.424 & -89.184 & 35.652 \\ 
        33100 & Miami & FL & 31201 & 70629 & -80.886 & 25.137 & -79.974 & 26.971 \\ 
        33340 & Milwaukee & WI & 9736 & 34310 & -88.542 & 42.842 & -87.069 & 43.544 \\ 
        33460 & Minneapolis & MN & 18333 & 57216 & -94.262 & 44.196 & -92.135 & 46.247 \\ 
        34980 & Nashville & TN & 11766 & 33716 & -87.567 & 35.408 & -85.779 & 36.652 \\ 
        35380 & New Orleans & LA & 8127 & 14178 & -90.964 & 28.855 & -88.758 & 30.712 \\ 
        35620 & New York & NY & 108292 & 313702 & -75.359 & 39.475 & -71.777 & 41.602 \\ 
        36420 & Oklahoma City & OK & 7638 & 18882 & -98.313 & 34.681 & -96.619 & 36.165 \\ 
        36740 & Orlando & FL & 14649 & 23406 & -81.958 & 27.642 & -80.861 & 29.277 \\ 
        37980 & Philadelphia & PA & 41170 & 124130 & -76.233 & 39.290 & -74.39 & 40.609 \\ 
        38060 & Phoenix & AZ & 18961 & 29693 & -113.335 & 32.501 & -110.448 & 34.048 \\ 
        38300 & Pittsburgh & PA & 14827 & 48812 & -80.519 & 39.721 & -78.974 & 41.173 \\ 
        38900 & Portland & OR & 10032 & 18974 & -123.786 & 44.886 & -121.514 & 46.389 \\ 
        39300 & Providence & RI & 9845 & 21788 & -71.907 & 41.096 & -70.752 & 42.096 \\ 
        39580 & Raleigh & NC & 7414 & 18822 & -78.995 & 35.255 & -78.007 & 36.266 \\ 
        40060 & Richmond & VA & 7636 & 24652 & -78.241 & 36.708 & -76.645 & 38.008 \\ 
        40140 & Riverside & CA & 21543 & 36396 & -117.803 & 33.426 & -114.131 & 35.809 \\ 
        40900 & Sacramento & CA & 8708 & 12620 & -122.422 & 38.018 & -119.877 & 39.316 \\ 
        41180 & Saint Louis & MO & 14772 & 49526 & -91.419 & 38.003 & -89.138 & 39.523 \\ 
        41620 & Salt Lake City & UT & 5336 & 11250 & -114.047 & 39.904 & -111.553 & 41.077 \\ 
        41700 & San Antonio & TX & 14943 & 43022 & -99.603 & 28.613 & -97.631 & 30.139 \\ 
        41740 & San Diego & CA & 14222 & 19166 & -117.611 & 32.529 & -116.081 & 33.505 \\ 
        41860 & San Francisco & CA & 20091 & 67850 & -123.174 & 37.054 & -121.469 & 38.321 \\ 
        41940 & San Jose & 6267 & CA & 6194 & -122.203 & 36.197 & -120.597 & 37.485 \\ 
        42660 & Seattle & WA & 17333 & 41986 & -122.853 & 46.728 & -120.907 & 48.299 \\ 
        45300 & Tampa & FL & 11974 & 19120 & -82.909 & 27.571 & -82.054 & 28.695 \\ 
        47260 & Virginia Beach & VA & 10405 & 29284 & -77.502 & 36.029 & -75.709 & 37.603 \\ 
        47900 & Washington & DC & 36271 & 100740 & -78.453 & 37.991 & -76.322 & 39.720 \\ 
\bottomrule
\end{tabular}
\caption{Summary of the 50 cities with their bounding box coordinates and size of the Twitter social network (nodes and edges).}
\label{tab:metro_information}
\end{table*}

\subsection{Sensitivity analysis of city-level regression}\label{sec:si:regression_sensitivity}

\begin{figure}[t!]
   \centering
      \includegraphics[width=0.99\linewidth]{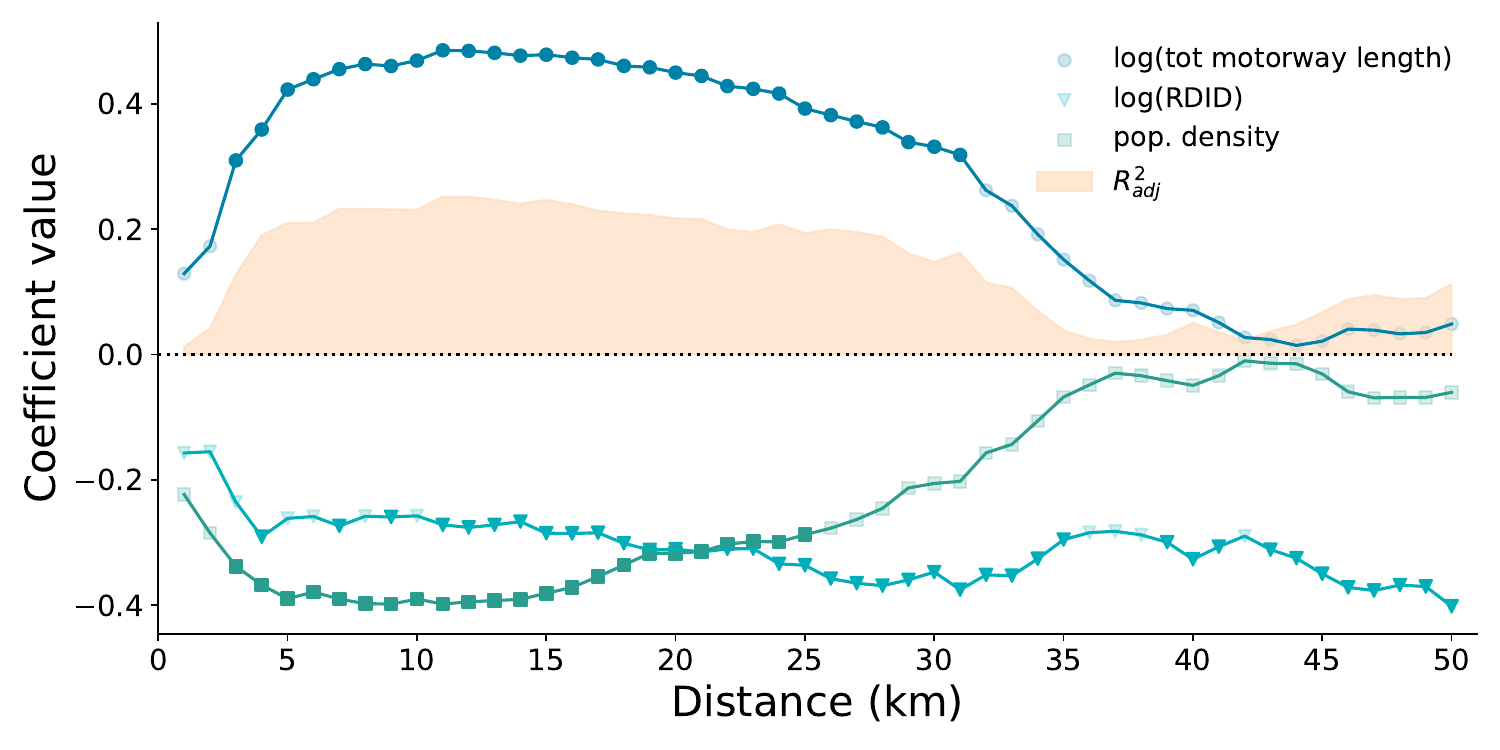}
      \caption{\textbf{Sensitivity analysis of city-level regression.} Value of $\beta$ coefficients and $R^2_{adj}$ for OLS regression models aimed at predicting the Barrier Score calculated considering only social ties of length up to $d$. Transparent bullets indicate non-significant coefficients.}
   \label{fig:regression_city_sensitivity}
\end{figure}

The regression results presented in Fig.~3 (main text) refer to a model which predicts a Barrier Score considering social ties with lengths up to 10 km ($B_D$, with $D=10km$). Fig.~\ref{fig:regression_city_sensitivity} shows the regression coefficients and adjusted $R^2$ for all values of $D$ ranging from 1km to 50km. The regression results hold in the range between 5km and 30km.

\subsection{Alternative regression models at the census tract pair level}
\label{sec:si:alt_regressions}

To test the robustness of the tract-level regression results, we first experiment with different ways of selecting the sample of tract pairs to include in the regression, as summarized in Table~\ref{tab:si_setting_explanation_tab}. The first three columns in Table~\ref{tab:si_regressions_tracts} compare different sampling criteria. Model (1) is our preferred OLS regression presented in the main text and serves as a benchmark. This model considers all tract pairs that are connected by at least one social tie (672,571 observations) or get connected by the rewiring process of our null model (1,996,095 observations). This is a natural choice of sampling, as it includes all pairs of tracts that can be potentially connected according to our null model, while excluding pairs of locations that are too sparsely populated and far apart to exhibit even a minimal level of social connectivity. Model (2) restricts the focus to tract pairs connected by at least one social tie observed in the real data, disregarding ties from the null model. Model (3) considers all possible pairs of tracts, regardless of whether social ties between them exist. This last sample contains a considerably higher number of pairs (almost 27M), the vast majority of which have no ties between them, neither in the real data nor in the null model.

The coefficient and significance levels are consistent across the three models, with the exception of the coefficient for the number of highways crossed, which turns to positive in model (3). Since the estimates for all other variables are stable, this could be attributed to adding many census tract pairs to our sample where there are no Twitter users at all, and hence no ties to other locations of the metropolitan area. These unconnected places are likely to be on the periphery of cities. The resulting coefficient on the number of highways crossed in this setting is meaningful in the sense that highways actually enhance accessibility and the likelihood of connections between such places. We address the changing role of highways by distance in detail in Section~\ref{sec:si:dist_barrier_interplots}. However, for our main models we rely on the sample behind model (1), as it allows more general interpretation, and because strength of connections within cities can only be meaningful in the context of sufficiently populated and sufficiently proximate locations.

\begin{figure}[t!]
   \centering
      \includegraphics[width=0.975\linewidth]{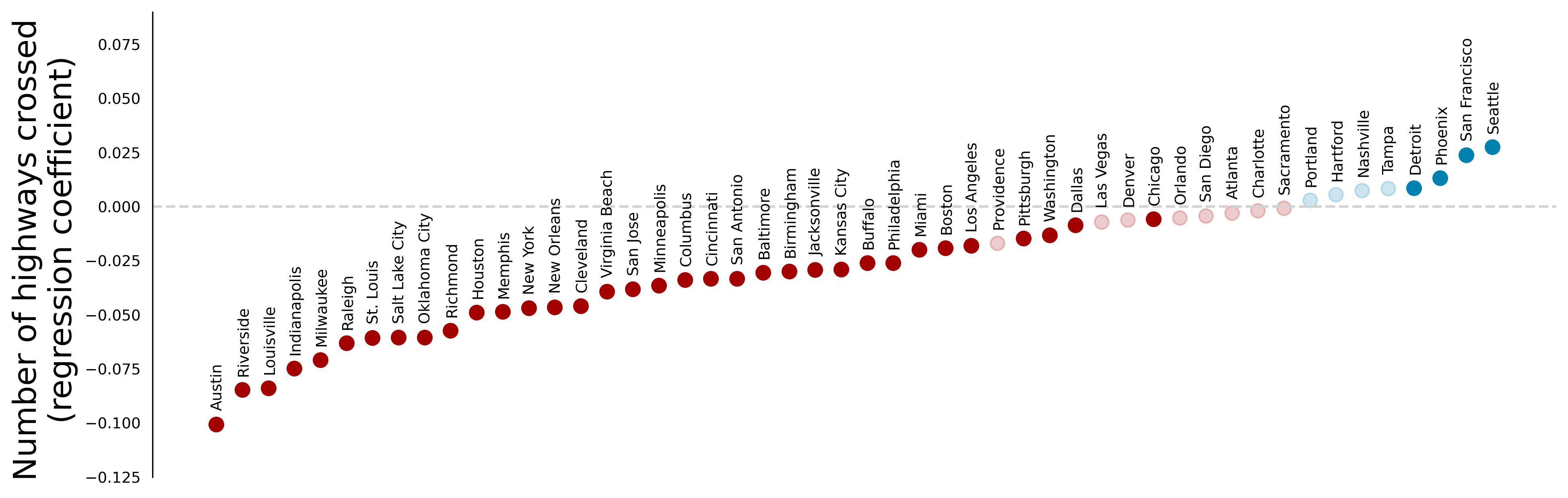}
      \caption{\textbf{Estimated effect of highways on social ties in each of the 50 metropolitan areas.} The coefficients are the results of separate models for each city, where the most detailed specification, Model (5) from Table~1 (main text) is used.}
   \label{fig:reg_coeff_for_each_metro}
\end{figure}

To further illustrate the robustness of our results, in Table~\ref{tab:si_regressions_tracts} we report two alternative models to our log-linear, OLS-based specifications using PPML (Poisson pseudo-maximum-likelihood) regressions. PPMLs expect a Poisson distribution and count data for the dependent variable, so they fit our data better. Model (4) based on our selected sample and model (5) based on the sample of all observed connections show identical results to OLS models in terms of sign an significance for all of our variables. Due to computational demand, it was not possible to fit PPML models to observations from all possible census tract pairs, but we expect results similar to the OLS setting. 

Our controlled correlations in Table~1 (main text) and all the above models are pooled regression models with fixed effects at the metropolitan area level. This means that census tract pair level observations from different cities are combined, and a dummy control variable is added to account for metropolitan area specificities. Here, we run model (5) from Fig.~1 (main text) for each of the 50 metropolitan areas separately, and report the coefficients for the number of highways crossed in Fig.~\ref{fig:reg_coeff_for_each_metro}. The results are in general consistent with the aggregated findings, with only 4 cities exhibiting positive and significant coefficients for that variable.

\begin{table}[t!]
\begin{center}
\begin{tabular}{@{\extracolsep{2pt}}lccc} 
& & & \\[-1.8ex] 
 & No null model ties & Null model ties & Total \\[-1.8ex] 
& & & \\ 
\hline \\[-1.8ex] 
 No social ties & 24,312,307 & \cellcolor{gray!25}1,996,095 & 26,308,402 \\[-1.8ex] 
& & & \\ 
Observed social ties & \cellcolor{gray!25}159,368 & \cellcolor{gray!25}513,203 & 672,571 \\[-1.8ex] 
& & & \\ 
\hline 
\end{tabular} 
\end{center}
\caption{\textbf{Sample composition behind our main models.} \textmd{Our null model is leveraged to construct the sample for our census tract pair level regressions. Light grey colors indicate the selected sample behind our preferred specifications.}}
\label{tab:si_setting_explanation_tab}
\end{table}

\begin{table*}[ht!]
\begin{center}
\begin{tabular}{@{\extracolsep{5pt}}lccccc} 
\hline \\[-1.8ex]
Estimator & OLS & OLS & OLS & PPML & PPML \\[-1.8ex]
& & & & & \\ 
Dependent variable & $log(1 + t_{ij})$ & $log(t_{ij} > 0)$ & $log(1 + t_{ij})$ & $t_{ij}$ & $t_{ij} > 0$ \\[-1.8ex]
& & & & & \\ 
& (1) & (2) & (3) & (4) & (5) \\[-2.5ex]
& & & & & \\ 
\hline \\[-1.8ex] 
 Nr highways crossed (log) & $-$0.021$^{***}$ & $-$0.043$^{***}$ & 0.010$^{***}$ & $-$0.383$^{***}$ & $-$0.186$^{***}$ \\ 
  & (0.001) & (0.001) & (0.003) & (0.007) & (0.005) \\[-1.8ex] 
  & & & & & \\ 
 Income abs difference & $-$0.018$^{***}$ & $-$0.006$^{***}$ & $-$0.002$^{*}$ & $-$0.154$^{***}$ & $-$0.001 \\
  & (0.0002) & (0.0004) & (0.001) & (0.003) & (0.002) \\[-1.8ex] 
  & & & & & \\ 
 Racial similarity & 0.027$^{***}$ & 0.021$^{***}$ & 0.003$^{***}$ & 0.336$^{***}$ & 0.083$^{***}$ \\ 
  & (0.0002) & (0.001) & (0.001) & (0.003) & (0.003) \\[-1.8ex]
  & & & & & \\ 
 Distance (log) & $-$0.086$^{***}$ & $-$0.138$^{***}$ & $-$0.034$^{***}$ & $-$1.291$^{***}$ & $-$0.611$^{***}$ \\ 
  & (0.0005) & (0.001) & (0.010) & (0.006) & (0.005) \\[-1.8ex] 
  & & & & & \\ 
 Population (product log) & 0.022$^{***}$ & 0.038$^{***}$ & 0.009$^{***}$ & 0.375$^{***}$ & 0.161$^{***}$ \\ 
  & (0.0004) & (0.001) & (0.001) & (0.012) & (0.004) \\[-1.8ex] 
  & & & & & \\ 
 Constant & 0.216$^{***}$ & 0.291$^{***}$ & 0.248$^{***}$ & $-$0.705$^{***}$ & 0.530$^{***}$ \\ 
  & (0.001) & (0.002) & (0.002) & (0.012) & (0.009) \\[-1.8ex] 
  & & & & & \\ 
\hline \\[-1.8ex] 
Metro fixed effect & Yes & Yes & Yes & Yes & Yes \\ 
Observations & 2,668,666 & 672,571 & 26,980,973 & 2,668,666 & 672,571 \\ 
\hline 
\end{tabular} 
\end{center}
\caption{Alternative regressions at the level of the census tract pairs to support our main models. \textmd{OLS stands for Ordinary Least Squares, while PPML stands for Poisson pseudo-maximum-likelihood. All independent variables are standardized in the same way for all models.}}
\label{tab:si_regressions_tracts}
\end{table*}

\begin{table*}[ht!]
\begin{center}
\begin{tabular}{@{\extracolsep{5pt}}lcccc} 
\hline \\[-1.8ex]
Estimator & OLS & OLS & OLS & OLS \\[-1.8ex]
& & \\ 
Dependent variable & $log(1 + t_{ij})$ & $log(1 + t_{ij})$ & $log(t_{ij} > 0)$ & $log(t_{ij} > 0)$\\[-1.8ex]
& & & & \\ 
& (1) & (2) & (3) & (4) \\[-1.8ex]
& &\\ 
\hline \\[-1.8ex] 
 Nr highways crossed (log) & $-$0.021$^{***}$ & $-$0.277$^{***}$ & $-$0.043$^{***}$ & $-$0.292$^{***}$ \\ 
  & (0.001) & (0.001) & (0.001) & (0.003) \\[-1.8ex]  
  & & & & \\
 Distance (log) & $-$0.086$^{***}$ & $-$0.244$^{***}$ & $-$0.138$^{***}$ & $-$0.296$^{***}$ \\ 
  & (0.0005) & (0.001) & (0.001) & (0.002) \\[-1.8ex]  
  & & & & \\ 
 Distance X Nr highways crossed &  & 0.196$^{***}$ &  & 0.201$^{***}$ \\ 
  &  & (0.001) &  & (0.002) \\[-1.8ex]  
  & & & & \\ 
 Income abs difference & $-$0.018$^{***}$ & $-$0.016$^{***}$ & $-$0.006$^{***}$ & $-$0.004$^{***}$ \\ 
  & (0.0002) & (0.0002) & (0.0004) & (0.0004) \\[-1.8ex]  
  & & & & \\ 
 Racial similarity & 0.027$^{***}$ & 0.025$^{***}$ & 0.021$^{***}$ & 0.019$^{***}$ \\ 
  & (0.0002) & (0.0002) & (0.001) & (0.001) \\[-1.8ex]  
  & & & & \\ 
 Population (product log) & 0.022$^{***}$ & 0.021$^{***}$ & 0.038$^{***}$ & 0.037$^{***}$ \\ 
  & (0.0004) & (0.0003) & (0.001) & (0.001) \\[-1.8ex]  
  & & & & \\ 
 Constant & 0.216$^{***}$ & 0.410$^{***}$ & 0.291$^{***}$ & 0.473$^{***}$ \\ 
  & (0.001) & (0.001) & (0.002) & (0.003) \\[-1.8ex]  
  & & & & \\ 
\hline \\[-1.8ex] 
Metro fixed effect & Yes & Yes & Yes & Yes \\ 
Observations & 2,668,666 & 2,668,666 & 672,571 & 672,571 \\ 
R$^{2}$ & 0.050 & 0.064 & 0.098 & 0.112 \\ 
\hline 
\end{tabular} 
\end{center}
\caption{Controlled correlations at the level of the census tract pairs including interaction terms.\textmd{The interaction of distance and number of highways crossed in model (2) and (4) are used to visualize the changing role of highways in the function of distance in Fig.~\ref{fig:si_interplots}.}}
\label{tab:si_interaction_model}
\end{table*}

\subsection{Decreasing Barrier Score with distance}\label{sec:si:dist_barrier_interplots}

Fig.~2 (main text) shows that the Barrier Score tends to decrease with distance. This suggests that the role of highways as barriers could change if more distant locations are considered. To quantify this decreasing barrier effect, we rely on our census tract level regression setting (see Table~1 in main text). First, we include the interaction effect of distance and number of highways crossed to our preferred regression specification. Interactions allow us to test conditional effects of one variable (in our case, distance) on the contribution of another variable (in our case, highways crossed) to the dependent variable (number of social ties). Model (1) in Table~\ref{tab:si_interaction_model} matches our final model from the main text, while Model (2) contains the interaction term. The interaction of distance and number of highways crossed is positive and significant, while the sign and significance of all other variables remain unchanged. Models (3) and (4) show the same results, but these models are based only on census tract pairs with observed social ties (see Table~\ref{tab:si_regressions_tracts}).

\begin{figure}[t!]
   \centering
      \includegraphics[width=0.99\linewidth]{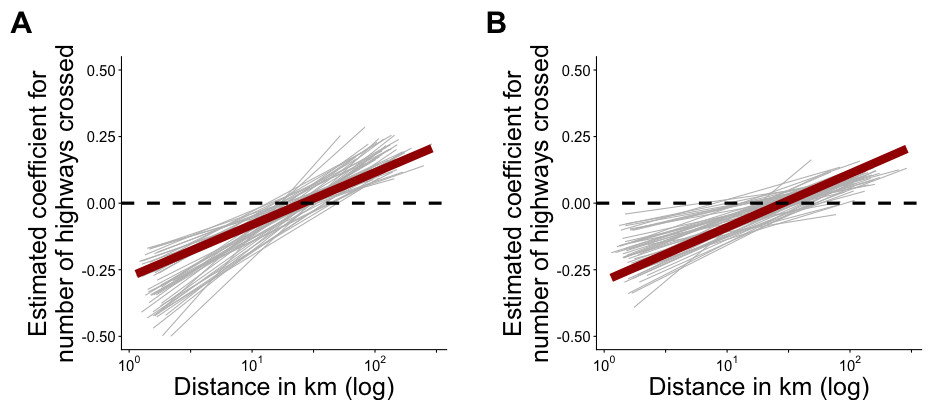}
      \caption{\textbf{The effect of highways on social ties at different distances.} \textbf{A.} is based on Model (2) of \ref{tab:si_interaction_model}, while \textbf{B.} is based on Model (4). Both figures suggests that a higher number of highways between pairs of tracts is associated with a lower number of social ties between them, up to an inter-tract distance of 20 kilometers. The red line illustrates the estimated effect and the associated confidence interval, which is very low for all distances in both model versions. The grey lines show the result of the same estimation for each city separately.}
   \label{fig:si_interplots}
\end{figure}

Second, we leverage this interaction term to visualize the changes in the coefficient of number of highways crossed in a two-way interaction term, conditional on the value of the other included variable, i.e.~distance of census tracts. In other words, here we plot the estimated effect of the number of highways crossed on the number of social connections at different distances. Figure~\ref{fig:si_interplots} shows that highways separating tracts within less than 20 km are associated with lower levels of social connectivity, whereas they are associated with a higher number of social ties for pairs of tracts that are farther apart. This suggests that highways might embody barriers to social ties at shorter distances, but they might foster accessibility (hence, opportunities for social connections) at longer distances.

\subsection{Barrier Scores of other street types}\label{sec:si:other_streettypes}

\begin{figure}[t!]
    \centering
    \includegraphics[width=\columnwidth]{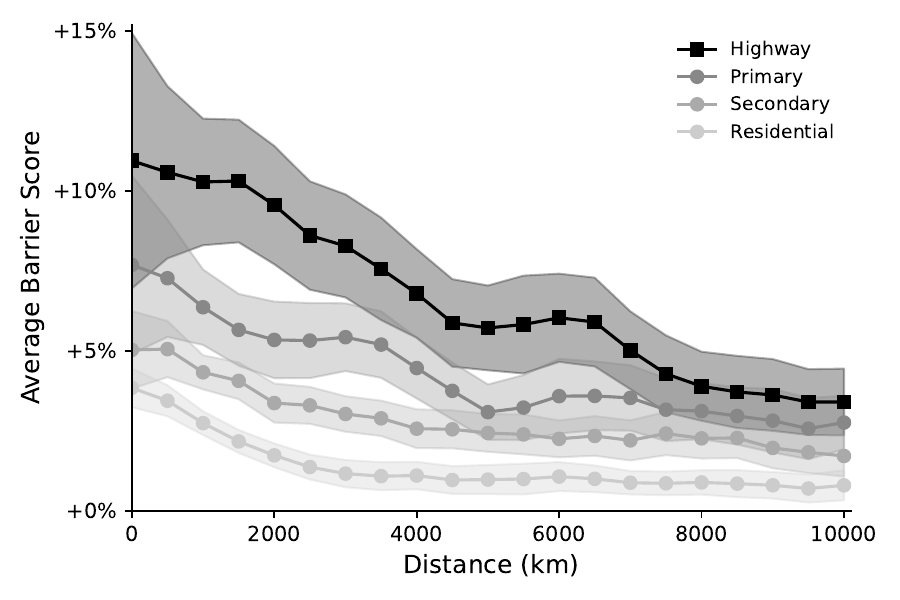}
    \caption{\textbf{The Barrier Score decreases with social tie distance for highways, to a lesser extent for other street types.} The distance-constrained Barrier Score $B(d)$ across multiple distances, averaged over all cities, and calculated for different types of roads. Across all distances, streets that are higher up in the road hierarchy have higher Barrier Scores.}
    \label{fig:streettypes}
\end{figure}

We compared the Barrier Score calculated on highways with Barrier Scores calculated considering other types of street included in Open Street Map's categorization. Fig.~\ref{fig:streettypes} presents the distance-constrained Barrier Score $B(d)$ for three further street types in addition to highways, by descending road hierarchy: \emph{primary} roads, \emph{secondary} roads, and \emph{residential} streets (i.e., streets that provide direct access to housing). While all road types yield positive scores, the highway score is markedly higher than all others.

\subsection{Randomization of highways}\label{sec:si:nullmodel_highways}

The observed decrease in Barrier Scores when considering streets with less vehicular traffic, such as residential streets, suggests that streets further down in the road hierarchy do not significantly impede mobility and social interactions. However, this interpretation could be confounded by the inverse correlation between the typical volume of vehicular traffic on streets of a given type and the abundance of that street type in the urban network. For instance, large American cities often feature only a small number of heavily trafficked highways, contrasted by a large number of residential streets. This raises the question of whether the observed Barrier Score patterns on highways are merely a statistical artifact of their relative scarcity.

To address this concern, we recompute the Barrier Scores on a hypothetical highway network of comparable length to the real network, but with randomly placed sections. To build the randomized version of a highway network of total length $L$, we use an iterative approach. At each iteration $i$, we select two random points within the bounding box of the target city, and found their respective closest nodes on the street network. We connect the two nodes with the shortest path between them, and add the obtained path to the randomized network. Let the length $L_i$ be the total length of the randomized network up to iteration $i$. To ensure a minimal residual between the real and randomized networks, we discarded any iteration $i$ that would cause the randomized network's size to exceed the real network size by more than 1\% (namely, when $L_{i} \geq 1.01 \cdot L$). The algorithm stops $L_i \approx L$, which we empirically represent with the range $[0.99 \cdot L \leq L_i \leq 1.01 \cdot L]$. The Barrier Scores are then averaged over 50 versions of the randomized layout for each city considered.

\begin{figure}[t!]
   \centering
      \includegraphics[width=0.99\linewidth]{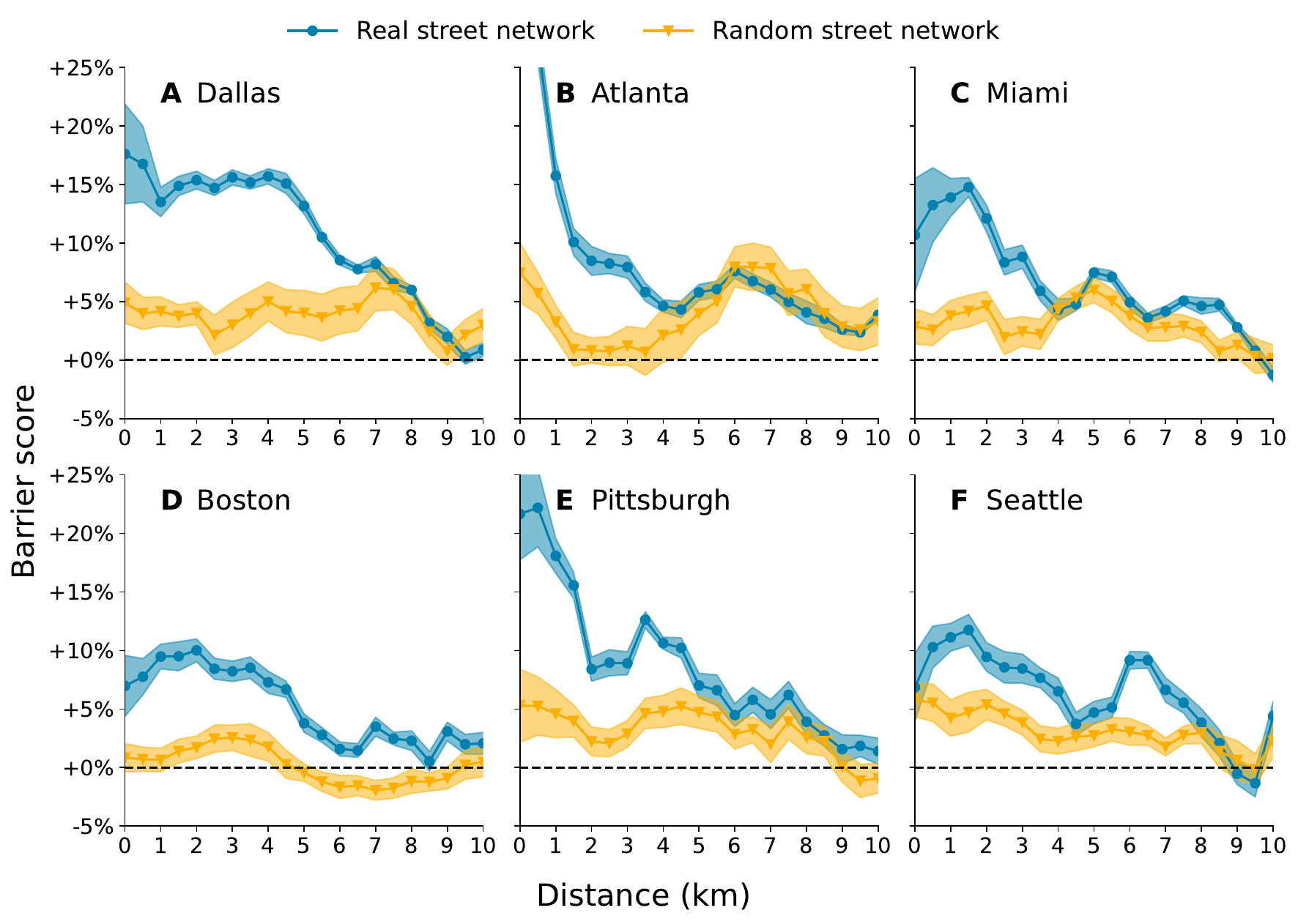}
      \caption{\textbf{Random street layout.} Barrier Score for social ties at distance $d$ for a model that considers the real highway network compared to a model that uses a randomized version of the highway network. Cities on top panels \textbf{A-C} are selected among cities without major natural barriers within their built environment, whereas cities on the bottom panels \textbf{D-F} are built around major water bodies.}
   \label{fig:random_street_layout}
\end{figure}

Fig.~\ref{fig:random_street_layout} presents the results for a selected set of cities. In all cases, the Barrier Scores  of the randomized street network were significantly lower than that of the real street network, particularly for ties of length up to 6-8km. This is true for cities that are not crossed by any major natural barriers (i.e., Dallas, Atlanta, Miami) as well as cities that are built around major water bodies (i.e., Boston, Pittsburgh, Seattle). However, the randomized street model still exhibits positive Barrier Scores, with values fluctuating with distance in a manner similar to the real highway network. This suggests that the randomized highway network is subject to the same spatial constraints imposed by the city's morphology and actual street network layout. For example, most shortest paths connecting the northern and southern parts of Seattle are bound to pass through the highway bridges that cross the Lake Union connecting Washington Lake to the bay (Interstate 5 or Route 99). Consequently, the spatial patterns of the randomized highway network cannot be fully disentangled from those of the real street network. Therefore, our results should be interpreted as upper bounds of the contribution of the highway network sparsity to the Barrier Score that we observe on the real data.

\subsection{Alternative social network data: Gowalla}\label{sec:si:gowalla}

\begin{figure}[t!]
   \centering
      \includegraphics[width=0.95\linewidth]{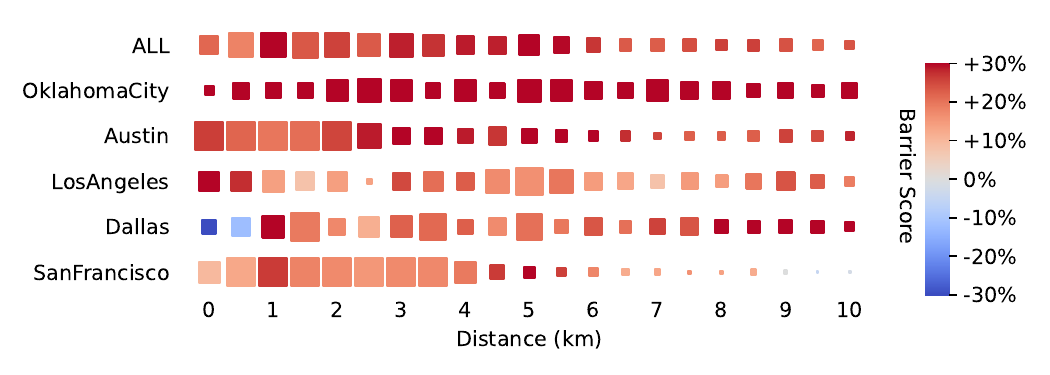}
      \caption{\textbf{Barrier Score vs. distance in the Gowalla database.} Heatmap of Barrier Scores $B(d)$ grouped into $0.5\,\mathrm{km}$ distance bands for the five cities most represented in the Gowalla dataset. Color denotes Barrier Score, areas of the squares denote relative number of links per distance band. The Barrier Scores are considerably higher than those observed on average in our Twitter dataset, across all distances.}
   \label{fig:gowalla}
\end{figure}

\begin{table}
\begin{center}
\begin{tabular}{ccc} 
\hline 
City & Users & Social links \\
\hline
Austin & 2,050  & 17,756 \\
Dallas & 1,539 &  8,054  \\
San Francisco & 1,254 &  5,070 \\
Los Angeles & 1,068 & 2,362 \\
Oklahoma City & 481  &  5,084 \\
\hline 
\end{tabular} 
\end{center}
\caption{Number of geo-referenced users and social ties between them in the five most represented cities in the Gowalla dataset}
\label{tab:si_gowalla_stats}
\end{table}

Studying the relationship between the built environment and social connections requires large-scale social network data with fine-grained geographical information. To illustrate the robustness of our results, we repeat our computations using the dataset from the online social platform Gowalla, derived from its API~\cite{si_cho_friendship_2011}. Gowalla is a location-based social network where users share their locations by means of check-ins. Unlike in Twitter, Gowalla is designed for connecting people who know each other in real life. The undirected friendship network from the site consist of 196,591 nodes and 950,327 edges. While the social network part of the data is similar to our Twitter network based on mutual followership, the geographic information available to determine the home location of users is different. 

The Gowalla data contains 6,442,890 check-ins across 50 cities performed by 107,092 users from February 2009 to October 2010. Inspired by~\cite{si_cho_friendship_2011}, we use the following procedure to determine the home location of users: First, as a minimum requirement, we focus on users who have at least 10 different visited locations in the dataset, and discard others. Second, we place check-ins into size 10 H3 hexagons (Uber's Hexagonal Hierarchical Spatial Index). These hexagons refer to an average 15.000 m2 area, which is close to the buffer area of a point with a 70 m radius. Third, we identify the hexagon in which each user made the most visits as their home location if they have at least 5 check-ins. We tested several more restrictive configurations, including time of day filters, but the results were the same for most users.

As a result of these filters, we were left with only five cities with a sufficient amount of data to extract reliable measurements (Table~\ref{tab:si_gowalla_stats}). The Barrier Score for these cities are considerably higher than those obtained from the Twitter data for the same cities~\ref{fig:gowalla}, which corroborates the validity of our results, and in addition suggests that the interplay between highways and social connection may be even more pronounced for stronger social ties.

\subsection{Beeline distance vs. walking distance}\label{sec:methods:nullmodel:walking}

\begin{figure}[t!]
   \centering
      \includegraphics[width=0.95\linewidth]{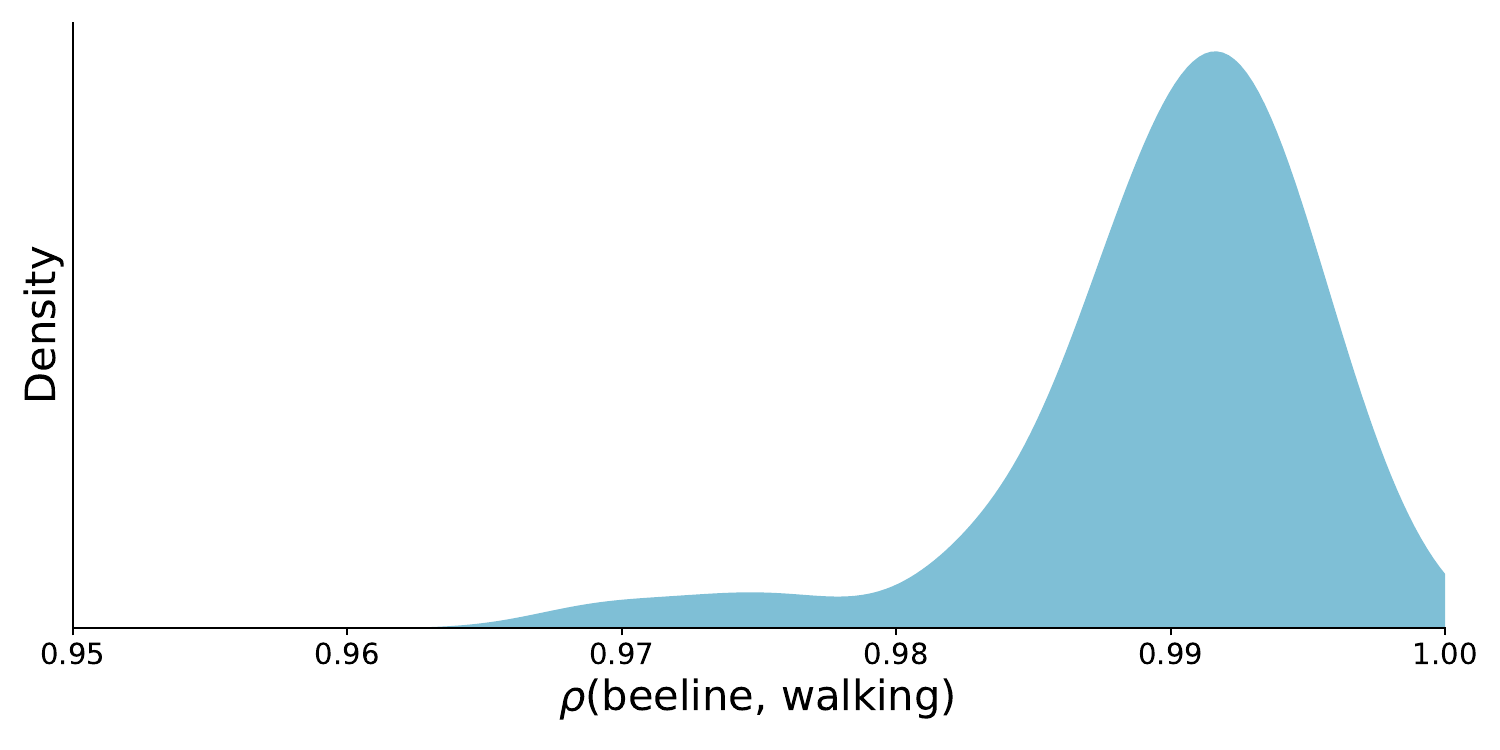}
      \caption{\textbf{Beeline distance vs. network distance.} Distribution (kernel density estimation) of the Pearson correlation coefficients between the length of social ties computed as the Euclidean distance between nodes and the length computed as the shortest path between them on the street network. We calculated 50 correlation values, one for each city. All correlations are very strong ($\rho>0.95$) and significant ($p<0.01$).}
   \label{fig:corr_beeline_walking}
\end{figure}

In our spatial modeling, we conceptualize social ties as straight segments connecting nodes, with the distance between points calculated as the length of these segments in Euclidean space, a measure often referred to as `beeline' distance. This approach facilitates an intuitive depiction of spatial distance and its efficient computation. To determine whether using straight segments is a good enough approximation, on a sample set of cities we find that results obtained using walking distance, defined as the length of the shortest path between two endpoints on the street network, are analogous to the results presented. This is expected, since beeline distance correlates very strongly with walking distance in all cities ($\rho > 0.95$, see Fig.~\ref{fig:corr_beeline_walking}).

\subsection{The Color of Highways: The racial context of US highway construction}
\label{sec:si:racialcontext}

Social segregation can happen along many different axes; one of the most obvious ones to scrutinize in contemporary US cities is race. Segregationist practices were ruled out by law only in the 1960s, with the Civil Rights Acts of 1964 and 1968, the former outlawing \textit{``discrimination or segregation on the ground of race, color, religion, or national origin''} \cite{si_general_records_of_the_united_states_government_civil_1964} and the latter explicitly expanding this principle to the provision (selling and renting) of housing~\cite{si_general_records_of_the_united_states_government_civil_1968}. Prior to this, however, explicitly racist and exclusionary urban policies such as redlining and racial zoning were widely practiced~\cite{si_rothstein_color_2017}, and their legacy is still lingering up until today~\cite{si_arcadi_concrete_2023}. The Interstate Highway System (IHS) is a compelling example thereof~\cite{si_schindler_architectural_2014, si_karas_highway_2015}. From the beginning of its construction in 1956, the IHS fostered urban sprawl for decades to come and played a major role in the suburbanization of US cities, which in turn is racially biased~\cite{si_rabin_highways_1973, si_boustan_was_2010, si_massey_suburbanization_2018, si_hadden_loh_separate_2020}. By the time that massive government-funded highway construction got underway throughout the country in the late 1950s, many US urban areas had majority Black inner cities. Many cities followed the advice of urban planner Robert Moses, not only on his proclaimed imperative that ``most of our new (...) expressways (...) must go right through cities and not around them''~\cite{si_moses_statement_1954}, but also in his unambiguously racialized suggestion that ``the practical solutions of the traffic problems in cities should be coordinated with slum clearance''~\cite{si_moses_statement_1954}, which meant building highways \textit{through} Black neighborhoods. The choice of specific sites within a city where highways would lead through sparked a great number of protests across the country, known as Freeway Revolts~\cite{si_mohl_stop_2004}. There are numerous, extensively studied examples of highway construction sites tearing apart or displacing historically Black communities~\cite{si_house_relocation_1970, si_howard_reframing_2016, si_howell_displacement_2020, si_mahajan_highways_2023}, proverbially known as \textit{``White roads through Black bedrooms''}~\cite{si_drummond_ayres_white_1967}. Likewise, many decision-makers used such major construction projects as an instrument to get rid of so-called ``ghettos'' or to spatially exclude underprivileged communities~\cite{si_sugrue_origins_2014, rothstein_color_2017, si_miller_roads_2018}. Effects of both practices are visible in today's patterns of racial residential segregation. Below, we explore these effects and their historical context for our nine case study cities.

\bigskip

\noindent\textbf{Cleveland, OH} has a long history of racial exclusion~\cite{si_yankey_neighborhood_2023}. Ever since urban highway construction gained momentum in the late 1950s, local infrastructural decisions have often been heatedly disputed at the intersection of race, class, and suburbanization~\cite{si_mohl_urban_2001, si_sisley_shaker_2017}. The two suburban neighborhoods of Shaker Heights and Cleveland Heights are well-known examples of these disputes. In the 1960s, when both neighborhoods were predominantly white and particularly wealthy, initial plans to construct a set of highways through these neighborhoods were successfully overturned. This sparked protests by local residents, supported by Carl Stokes, the first Black mayor of a major US city~\cite{si_dawson_saving_2023, si_doll_clevelands_2024}. From 1976 onwards, traffic diverters were installed at the suburban fringe of Shaker Heights, dubbed by some as the ``Berlin Wall'' for Black people~\cite{si_schindler_architectural_2014} and de facto keeping out a lower income minority; many inhabitants saw the traffic diversion measures as racially connotated~\cite{si_schindler_architectural_2014, si_sisley_shaker_2017}. As of today, both Shaker Heights and Cleveland Heights are regarded by some as a symbol of successful suburban racial integration~\cite{si_souther_through_2023}. Overall, however, Cleveland is one of the poorest and most racially segregated among major US cities~\cite{si_berube_metro_2019, si_menendian_roots_2021, si_yankey_neighborhood_2023}. At the same time, Cleveland ranks highest in our computations of the city-wide Barrier Score. Moreover, the Barrier Score is high for I-77, which separates the cluster of predominantly Black communities in the east from the rest of the metropolitan area (see Fig.~4\textbf{A} in the main text). 

\bigskip

\noindent\textbf{Orlando, FL.} In 1961, the I-4 in Florida became one of the first highways supported by the Federal-Aid Highway Act of 1956 to officially celebrate its opening~\cite{si_federal_highway_administration_greatest_2017}. Within Orlando, the I-4 was built parallel to Division Street (today Division Avenue), a ``dividing line''~\cite{si_brotemarkle_crossing_2006} between white and Black neighborhoods. The I-4 further emphasized this racial division~\cite{si_brotemarkle_crossing_2006, si_wolf_new_2018}. The neighborhood of Parramore, once a thriving Black community, suffered a particularly heavy impact from the construction, with 551 Black properties displaced. I-4 thus literally cemented an already existing racial separation, further acting as a  ``class barrier''~\cite{si_gama_parramore_2015} as it separated Parramore from downtown Orlando~\cite{si_sherman_connections_2022}. The construction of the Expressway 408 further exacerbated this impact, isolating the Griffin Park public housing project from the rest of Parramore~\cite{si_gama_parramore_2015}. In a 2006 report, the City of Orlando identified the 408 as a ``development barrier''~\cite{si_city_of_orlando_downtown_2006} for the neighborhood. As of today, Orlando still remains highly segregated, with the I-4 cutting through the city ``like a picket line''~\cite{si_wolf_new_2018}. At the same time, both I-4 and Expressway 408 have high Barrier Scores in our results, as shown in Fig.~4\textbf{B} in the main text. 

\bigskip

\noindent\textbf{Milwaukee, WI.} Next to Cleveland, Milwaukee is the second city which appears both in the list of 10 most segregated US cities \cite{si_othering__belonging_institute_most_2019} and in the top 3 of our Barrier Score ranking (Fig.~4\textbf{C} in the main text). In the 1960s, when the Black population of Milwaukee was still forcefully constrained to live in the North Side, the city was home to numerous protests against segregation. In 1967, for example, the Milwaukee’s NAACP (National Association for the Advancement of Colored People) Youth Council marched from North to South across the 16th Street Bridge, which was jokingly called ``the longest bridge in the world'' for connecting Africa and Poland, given its location between the majority-Black North Side, and the (at that time) almost exclusively Polish Old South Side~\cite{si_wisconsin_historical_society_crossing_2024}. The construction of I-43 in the 1960s severely impacted the North Side’s Bronzeville neighborhood, a formerly vibrant Black community. Many of Bronzeville’s inhabitants were displaced, and commercial areas were demolished~\cite{si_house_relocation_1970, si_mcbride_first_2007}. I-43 also impacted the South Side, leading to housing shortage. However, while other communities within Milwaukee were dissipated as a consequence of highway construction, the Southside remained ``solidly Polish''~\cite{si_lackey_milwaukees_2013}. In our computations for Milwaukee, we see that both the I-794 (separating the majority Black North from the majority white South), and the I-43 (the northern part of which disrupted many Black neighborhoods) have high Barrier Scores, as illustrated in Fig.~4\textbf{C} in the main text.

\bigskip

\noindent\textbf{Oklahoma City, OK.} In contemporary Oklahoma City, the ``city's divided soul''~\cite{si_felder_citys_2014} is most prominently expressed in the I-235, which separates majority Black neighborhoods in the East from majority white neighborhoods in the West. In contrast to previously mentioned cities, however, urban highway construction in Oklahoma City did not appear as immediate impact on previously thriving communities. For example, the I-235 (formerly known as the Centennial Expressway) was constructed in the late 1980s, creating a link between I-35, I-40, and I-44, and simultaneously isolating the largest historically Black neighborhood of Deep Deuce from the rest of the inner city~\cite{si_welge_oklahoma_2007}. However, the construction of I-235 was not so much a catalyst of Black neighborhood destruction, but rather the final blow to a neighborhood whose vitality and street life had already been eroded by several decades of car-centric planning and ``urban renewal''~\cite{si_payne_newbuild_2019}. As civil right activist James Baldwin succinctly put it in an interview with Kenneth Clark in 1963, ``Urban renewal (...) means Negro removal''~\cite{si_baldwin_conversation_1963}. Under the auspices of OCURA (Oklahoma City Urban Renewal Authority), established in 1961, entire neighborhoods of Oklahoma City fell victim to major construction projects~\cite{si_oklahoma_city_urban_renewal_authority_annual_1963, si_lackmeyer_historically_2021}. The University Medical Center urban renewal project, for example, displaced over 700 families, over 90\% of which were Black~\cite{si_digital_scholarship_lab_family_2024}. Lastly, in the recent two decades, Oklahoma City has witnessed a complex process of gentrification, most prominently underway in Deep Deuce~\cite{si_tierney_gentrification_2015, si_payne_newbuild_2019}. The Barrier Scores we computed for Oklahoma City are particularly high in the inner city, for all highways mentioned above: I-35, I-40, I-44, and I-235 (see Fig.~4\textbf{D} in the main text).

\bigskip

\noindent\textbf{Detroit, MI} is known to be not only within the top 10 metropolitan areas by numbers of Black inhabitants \cite{si_moslimani_facts_2024}, but also as one of the most racially segregated cities in the US~\cite{si_othering__belonging_institute_city_2024}. The city has a complex history in which the rise of the automobile, then deindustrialization, suburbanization and political marginalization are intertwined with a legacy of systemic and physical violence against Black people~\cite{si_sugrue_origins_2014}. The Birwood Wall (also known as Eight Mile Wall or Wailing Wall) is an infamous concrete symbol of Detroit’s predicament. It was built in 1941 by a private developer, with the aim of securing governmental funds for the construction of an all-white residential complex, which, as by the Federal Housing Administration's requirements, had to be physically separated from the adjacent majority Black area redlined as ``slum''~\cite{si_sugrue_origins_2014, si_einhorn_segregation_2021}. As of today, the Birwood Wall is still standing, located next to the Eight Mile Road, which in turn is closely associated with racial segregation in popular culture, manifesting the divide between the Black inner city and the white suburbs~\cite{si_michigan_radio_newsroom_8_2014}. From the late 1940s onwards, land clearing for highway construction additionally exacerbated Detroit’s already ongoing housing crisis, displacing tens of thousands of residents, most of them Black, often on short notice and without proper assistance to find a new dwelling~\cite{si_sugrue_origins_2014, si_whitaker_understanding_2023}. Numerous Black neighborhoods were destroyed by highway construction: the I-75 and I-375 practically erased Black Bottom, Hastings Street, Paradise Valley, and parts of the Lower East Side; while the formerly coherent urban fabric in the Western part of the city got bisected by I-94 and M10~\cite{si_da_via_brief_2012, si_sugrue_origins_2014, si_miller_roads_2018, si_whitaker_understanding_2023}. For all highways mentioned above, the Barrier Scores in our computations are at their highest within the city center, in proximity of the historically Black (former) neighborhoods of Black Bottom and Paradise Valley, while Eight Mile Road stands out as a city limit delineation (marking the transition between counties) with a particularly high Barrier Score (see Fig.~4\textbf{E} in the main text). 

\bigskip

\noindent\textbf{Austin, TX.} As of today, Austin is the most economically segregated large metropolitan area in the US~\cite{si_florida_segregated_2015}. At the same time, as the Black Austin Coalition underlines, income inequality amongst Austin’s residents is strongly associated with race~\cite{si_woods_black_2012}. The infamous Austin City Plan of 1928 suggested to ``solve'' the ``race segregation problem'' (the ``problem'' being \textit{how} to segregate Black people within ``constitutional'' limits) by implementing a set of measures which would force Black people into moving to the East side of the city, with the East Avenue as officially proposed segregation line. The implementation of the City Plan had the foreseen consequences of Black Austinites being forced to move to the East side of the city~\cite{si_skop_austin_2010, si_woods_black_2012}. Thus, thirty years later, building the I-35 along East Avenue meant ``solidifying the dividing line''~\cite{si_reconnect_austin_history_2023} between two racially disparate parts of the city, notably with only a handful of crossings between East and West. Our Barrier Score computations also depict the I-35 as clear dividing line between two parts of the city (Fig.~4\textbf{F} in the main text). Over the following decades, a combination of systemic neglect, ``urban renewal'' (including the seizing of Black property and systematic funding of non-Black serving projects), and suburban sprawl fostered by highway presence severely impacted Black communities in East Austin~\cite{si_woods_black_2012, si_bernier_highway_2023, si_charpentier_two_2023}. In the present time, gentrification is underway in East Austin, with real estate prices rising tenfold in the last 20 years. Simultaneously, Austin is currently the only large city in the US that is suffering a net decrease in Black inhabitants \textit{in spite of} an overall rapidly growing population~\cite{si_buchele_austins_2016}. As of 2024, plans are underway for the I-35 to be expanded from 16 to 22 lanes in Downtown~\cite{si_bernier_your_2024}. 

\bigskip

\noindent\textbf{Columbus, OH} Several highways (I-70, I-71, I-670) bisect the urban fabric of Columbus, Ohio (see Fig.~4\textbf{G} in the main text). The alignment of these highways with former redlined areas is particularly startling~\cite{si_nelson_mapping_2023}. The chilling words of Warren Cremean, an official of Ohio's Department of Transport, illustrate the intentionality behind routing decisions: ``[W]e married highway money and urban renewal money and wiped out (...) the worst slum in the state of Ohio'' (quoted in an interview with Rose and Seely (1990)~\cite{si_rose_getting_1990}). In several locations in the city, highway trajectories went directly through redlined neighbourhoods. Flytown, a diverse, but socioeconomically disadvantaged neighbourhood with many Black and Irish inhabitants, was erased from the map to make way for I-670~\cite{si_smith_african-american_2014}. Large parts of the Near East Side, which at the time of construction was predominantly Black, faced large-scale destruction by all three major highways in the city: I-70, I-71 and I-670. Two historically Black neighbourhoods in the Near East Side were particularly affected. I-70 was built right through Hanford Village, by that time a bustling Black community, destroying many homes and cutting of the Western part of the neighbourhood from the rest of it~\cite{si_smith_african-american_2014}. Similarly, King-Lincoln/Bronzeville saw the construction of I-71 through its core, entailing the demolition of homes and businesses and severely hindering access to this Black community in combination with I-670 along its borders; at the same time, the predominantly white, affluent neighborhood Bexley, east of Bronzeville, was spared from the highways~\cite{si_thompson_how_2020}. On a larger scale, I-71 disconnected the predominantly Black Near East Side from Downtown. 

\bigskip

\noindent \textbf{Richmond, VA} was the Capital of the Confederacy during the Civil War; in the following decades, the city became deeply segregated, with redlining practices further exacerbating racial divisions~\cite{si_nelson_mapping_2023}. The neighbourhood of Jackson Ward gained key importance for Richmond’s Black community, dubbed ``Black Wall Street'' and ``Harlem of the South'' prior to World War II. In the 1960s, however, the I-95 and the I-64/I-95 interchange were built directly through Jackson Ward, destroying numerous homes and businesses, and cutting off the neighbourhood’s northern part, Gilpin (see Fig.~4\textbf{H} in the main text). The public housing community of Gilpin Court was thus cut off both from Jackson Ward and from the rest of Richmond, with only a few physical connections bridging across the highway. ``Urban renewal'' brought further displacement to Jackson Ward through large-scale construction projects such as the Coliseum~\cite{si_howell_displacement_2020}. Population decline, urban neglect and continuously high poverty rates further impacted Richmond communities in the vicinity of the newly constructed highways, particularly north of I-95~\cite{si_howard_reframing_2016, si_richmond_redevelopment_and_housing_authority_rrha_jackson_2023}. Historically, Richmond's segregationist policies in public housing further contributed to a ``concentration of racialized poverty''~\cite{si_howard_reframing_2016}. Today, Richmond's inhabitants are confronted with particularly high eviction rates entangled with racialized dispossession~\cite{si_howell_displacement_2020}, giving rise to initiatives like Residents of Public Housing in Richmond Against Mass Evictions (RePHRAME)~\cite{si_rephrame_residents_2014}. Most recently (as of 2023), the city of Richmond has obtained a grant from the US department of transportation, dedicated to ``reconnecting'' Jackson Ward with the rest of the city across highway division lines~\cite{si_reconnect_jackson_ward_reconnect_2023}. 

\bigskip

\noindent \textbf{Nashville, TN.} Nashville's urban landscape is heavily fragmented by three major interstates: I-24, I-40, and I-65 (see Fig.~4\textbf{I} in the main text). The trajectory of I-40 displays a ``kink in the road''~\cite{si_haynes_one_2020} through North Nashville: When going from West to East, instead of following Charlotte Avenue, I-40 is routed one mile further north. The construction of I-40 began only in 1967; its exact route, however, had already been decided a decade prior, without duly involving the public. Initial plans with a more direct routing of I-40 had then been discarded in favor of introducing said ``kink in the road'', which implied major disruptions for the majority Black North Nashville. Due to misleading information and legal process violations from the authorities’ side, the actual plans for I-40 and the destruction that it entailed only became clear to the residents of North Nashville as construction had already begun~\cite{si_haynes_one_2020}. In response, the Nashville I-40 Steering Committee was formed; their case \textit{Nashville I-40 Steering Committee v. Ellington} won a temporary restraining order –- the first time that highway construction had been ``halted by claims of racial discrimination''~\cite{si_mohl_interstates_2002}. However, the case was ultimately lost in federal court; the I-40 ended up ripping through North Nashville as planned and ``bulldozed local prosperity in the name of national economic development''~\cite{si_arcadi_concrete_2023}. I-40 
was constructed along and through Jefferson street, Nashville’s main Black business and cultural district, which became severely impacted and divided. 80\% of Black-owned businesses were either directly demolished, or damaged through reduced accessibility for clients. The same highway also ``cut in half a thriving academic cluster''~\cite{si_haynes_one_2020} as it separated three major Black higher education institutions, Fisk University, Meharry Medical College, and Tennessee A.~\&~I.~University (later Tennessee State University), both from each other \textit{and} from surrounding majority Black neighbourhoods. Real estate values dropped in the area and housing conditions quickly and severely deteriorated~\cite{si_haynes_one_2020}. Ultimately, the construction of I-40 and its consequences represent a decisive contribution to today's rampant poverty rates in the area~\cite{si_perry_restore_2021}.

\end{document}